\pdfoutput=1
\documentclass[aps,prd,reprint,superscriptaddress,longbibliography,nofootinbib]{revtex4-1}

\usepackage{graphicx}
\usepackage{amsmath}
\usepackage{amssymb}
\usepackage{amsfonts}
\usepackage{dcolumn}
\usepackage{dsfont}
\usepackage{latexsym}
\usepackage{rotating}
\usepackage{color}
\usepackage{latexsym}
\usepackage{bbm}
\usepackage{subfigure}
\usepackage{float}
\usepackage{epsfig}
\usepackage{psfrag}
\usepackage{natbib, hyperref}
\usepackage{bm}
\usepackage{amsthm}
\usepackage{eucal}
\usepackage{mathrsfs}
\usepackage{url}
\usepackage{braket}
\usepackage{ulem}
\usepackage{calligra}
\usepackage[T1]{fontenc}
\usepackage[utf8]{inputenc}
\usepackage{newunicodechar}
\usepackage{marvosym}
\usepackage[absolute]{textpos}
\usepackage[cal=cm]{mathalfa}
\usepackage{soul}

\usepackage{tcolorbox}

\usepackage{color}

\usepackage{hyperref}
\hypersetup{
colorlinks=true,final=true,
        linkcolor=blue,
        citecolor=blue,
        filecolor=blue,
        urlcolor=blue,
}

\begin{document}
\setlength{\abovedisplayskip}{3pt}
\setlength{\belowdisplayskip}{3pt}

\title{Skyrmion Control of Majorana States in Planar Josephson Junctions}
\thanks{\tiny This manuscript has been authored by UTBattelle, LLC under Contract No. DE-AC05-00OR22725 with the U.S. Department of Energy. The United States Government retains and the publisher, by accepting the article for publication, acknowledges that the United States Government retains a non-exclusive, paid-up, irrevocable, world-wide license to publish or reproduce the published form of this manuscript, or allow others to do so, for United States Government purposes. The Department of Energy will provide public access to these results of federally sponsored research in accordance with the DOE Public Access Plan (http://energy.gov/downloads/doepublic-access-plan)}
\author{Narayan Mohanta}
\affiliation{Materials Science and Technology Division, Oak Ridge National Laboratory, Oak Ridge, TN 37831, USA}
\author{Satoshi Okamoto}
\affiliation{Materials Science and Technology Division, Oak Ridge National Laboratory, Oak Ridge, TN 37831, USA}
\author{Elbio Dagotto}
\affiliation{Materials Science and Technology Division, Oak Ridge National Laboratory, Oak Ridge, TN 37831, USA}
\affiliation{Department of Physics and Astronomy, The University of Tennessee, Knoxville, TN 37996, USA
}

\begin{abstract}
Planar Josephson junctions provide a versatile platform, alternative to the nanowire-based geometry, for the generation of the Majorana bound states, due to the additional phase tunability of the topological superconductivity. The proximity induction of chiral magnetism and superconductivity in a two-dimensional electron gas showed remarkable promises to manipulate topological superconductivity. Here, we consider a  Josephson junction involving a skyrmion crystal and show that the chiral magnetism of the skyrmions can create and control the Majorana bound states without the requirement of an intrinsic Rashba spin-orbit coupling. Interestingly, the Majorana bound states in our geometry are realized robustly at zero phase difference at the junction. The skyrmion radius, being externally tunable by a magnetic field or a magnetic anisotropy, brings a unique control feature for the Majorana bound states.
\end{abstract}

\maketitle

\noindent The unification of non-trivial spin texture and superconductivity via advanced interface engineering is a futuristic approach to create and manipulate non-Abelian Majorana bound states (MBS) for their controlled usage in fault-tolerant topological quantum computing~\cite{Kitaev_AnnPhys2003,Desjardins_NatMat2019,Yazdani_NMat2019,Nayak_RMP2008,Alicea_NatPhy2011}. The nanoscale control of magnetism not only relaxes the need for a specific form of Rashba spin-orbit coupling, but also motivates for a magnetic field-free platform for the braiding of the MBS~\cite{Kjaergaard_PRB2012,Klinovaja_PRL2013,Klinovaja_PRX2013,Geoffrey_PRL2016,Mohanta_PRApplied2019,Tong_PRB2019,Elbio_arxiv2020}. Despite numerous successes in the search for the MBS in one-dimensional geometries, the associated limitations such as the intrinsic instabilities of one-dimensional systems, the need for fine tuning of parameters, and the technological obstacles in physical implementation, suggest to look for a two-dimensional platform~\cite{Shabani_PRB2016,Karzig_PRB2017}. The discovery of topological superconductivity in phase-controlled planar Josephson junctions is, therefore, a major step towards the realization of a two-dimensional array of MBS for designing scalable braiding protocols~\cite{Pientka_PRX2017,Ren_Nature2019,Fornieri_Nature2019,Javad_PRL2021}. The Josephson junction geometry provides additional control to tune the MBS by changing the shape of the junction, strain and unconventional spin-orbit coupling~\cite{Tong_PRL2020, Alidoust_PRB2018,Alidoust_PRB2021,Setiawan_PRB2019_1,Alex_PRRes2021}. A time-reversal invariant topological superconductivity can also be induced by placing the Josephson junction on top of a strong topological insulator~\cite{Fu_PRL2008}. Previous works on the Josephson junction-based platforms, however, revealed the requirements of a strong intrinsic Rashba spin-orbit coupling and $\pi$-phase biasing of the Josephson junction. These constraints pose serious challenges in the detection and manipulation of the MBS  under realistic conditions. Chiral magnetism in proximity to an $s$-wave superconductor generates exotic effects including the appearance of the Majorana modes~\cite{Loss_PRB2016,Kovalev_PRB2018,Garnier_CommunPhys2019,Mohanta_PRB2020,Mascot_npjQM2021,Bedow_PRB2020,Rex_PRB2019,Rex_PRB2020}; however, the location and stability of the Majorana states in these platforms are difficult to anticipate due to their non-localized nature.

\begin{figure}[t]
\begin{center}
\epsfig{file=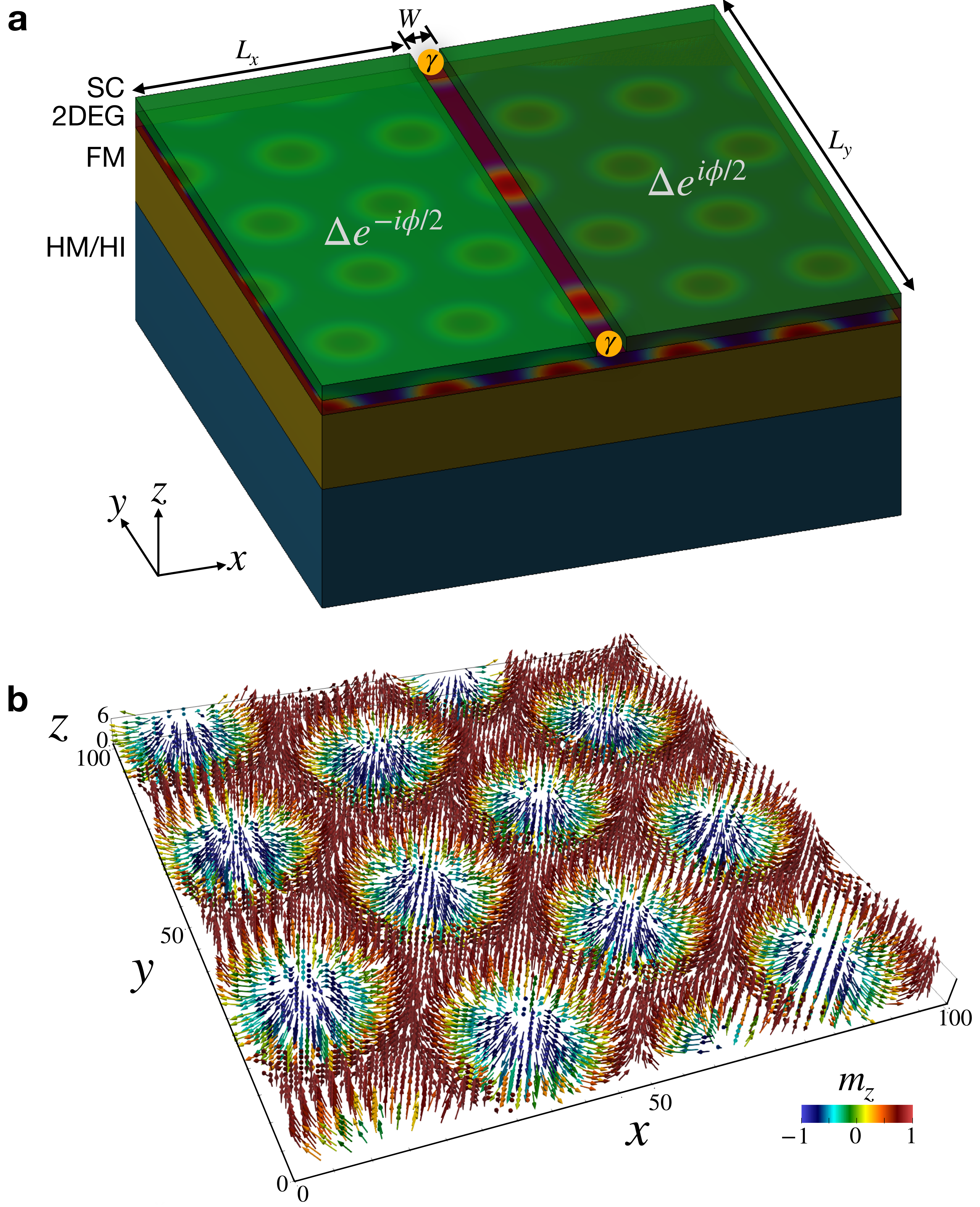,trim=0.0in 0.0in 0.0in 0.0in,clip=false, width=84mm}
\caption{{\bf Device geometry and a skyrmion crystal.}  {\bf a} Planar Josephson junction on top of a skyrmion crystal. The two-dimensional electron gas (2DEG) exhibits both proximity-induced superconductivity from the top superconductor (SC) layers and spatially-varying magnetism from the bottom skyrmion crystal that is created in the ferromagnet due to the competition between  exchange interactions in the ferromagnet (FM) and the heavy metal or heavy insulator (HM/HI), with a field or anisotropy. The zero-energy Majorana bound states (shown as yellow bubbles) are localized at the two ends of the quasi-one-dimensional metallic channel. {\bf b} The skyrmion crystal spin texture, spontaneously generated in a Monte Carlo simulation using a 100$\times$100$\times$6 lattice with ferromagnetic exchange interaction strength $J\!=\!1$, Dzyaloshinskii-Moriya interaction strength $D\!=\!0.3J$, magnetic field $H_z\!=\!0.1J$, spin amplitude $S\!=\!1$, and easy-plane anisotropy $A=0.01J$. The colorbar in {\bf b} denotes the $z$ component of the magnetization $m_z$.}
\label{fig1}
\vspace{-9mm}
\end{center}
\end{figure}

In our considered geometry, the planar Josephson junction, composed of a two-dimensional electron gas and an $s$-wave superconductor, is placed on top of a N\'eel-type skyrmion crystal (SkX) in such a way that the two-dimensional electron gas experiences the spatially-varying magnetic field from the bottom SkX and  it is also proximitized to the electron pairing from the top superconductors, as described in Fig.~\ref{fig1}{\bf a}. The interplay between the SkX spin texture and the proximity-induced superconductivity leads to topological superconductivity near the middle quasi-one-dimensional channel of the Josephson junction with localized MBS at its two ends. The advantages of using the SkX are: (i) the chiral magnetism generates a robust fictitious gauge field, that can also be visualized as a spin-orbit coupling, and a local Zeeman field which remove the stringent criteria of a strong Rashba-type spin-orbit coupling and, therefore, essentially expands the region of parameter space to realize the MBS, (ii) the existence of the MBS can be further controlled externally by tuning the skyrmion radius, and (iii) usual planar Josephson junctions are required to be phase biased with a phase difference $\varphi \!=\! \pi$, between the two superconducting regions, to minimize the critical magnetic field for the topological transition and to maximize the chemical-potential range within which the MBS appear~\cite{Pientka_PRX2017}; the current SkX-based Josephson junction is not required to be phase biased and the MBS can be found robustly at $\varphi \!=\! 0$. Using the zero-energy feature of the quasiparticle states with a topological energy gap, sharp localization of these states, charge-neutrality condition, two order parameters, {\it viz.} Majorana polarization and curvature of the density of states, we confirm the existence of the MBS in our set up. The tunable phase difference and the skyrmion radius together provide broad, flexible control of the MBS which is indispensable to achieve the long-sought-after goal of the braiding.\\

\vspace{1em}
\noindent \textbf{Results\\}
{\bf Theoretical set up.} For the generation of the SkX, we consider a heterointerface of a thin-layer ferromagnet and a heavy compound (metal or insulator). The advantage of the heavy compound is that it helps to generate a large Dzyaloshinskii-Moriya interaction (DMI) at the interface between the ferromagnet and the heavy compound. The cooperation between the DMI and the ferromagnetic exchange interaction of the ferromagnet produces a triangular SkX, in the presence of a magnetic field or an anisotropy. Our Monte Carlo simulations reveal that columns of skyrmions, arranged in a triangular array, appear spontaneously within a six-layer ferromagnet, although the DMI exists predominantly at the interface between the ferromagnet and the heavy compound, as shown in Fig.~\ref{fig1}{\bf b}. We perform simulated annealing using the Metropolis energy-minimization algorithm, formulated with the following Hamiltonian
\begin{align}
\mathcal{H}\!=\!&-J\sum_{\langle ij \rangle} \mathbf{S}_i \cdot \mathbf{S}_j -D \sum_{\langle ab \rangle}(\hat{z} \times \hat{r}_{ab}) \cdot ( \mathbf{S}_a \times \mathbf{S}_b  ) \nonumber \\
&-H_z\sum_{i}S_{zi}-A\sum_{a}|S_{za}|^2,
\label{Ham}
\end{align}
where $J$ is the nearest-neighbor ferromagnetic exchange interaction strength in the ferromagnet, $D$ is the DMI strength at the bottom ferromagnet layer that interfaces with the heavy compound, $H_z$ is the perpendicular magnetic field, $A$ is the easy-plane magnetic anisotropy at the bottom ferromagnet layer, $i$,$j$ are the site indices in the entire ferromagnet, and $a$,$b$ are the two-dimensional site indices at the bottom ferromagnet layer. The DMI, present dominantly at the interface between the ferromagnet and the heavy compound, generates a N\'eel-type SkX~\cite{NM_PRB2019,NM_PRB2020,Mohanta_CommunPhys2020}. Besides the  engineered interfaces, the SkX naturally appears in a wide variety of materials~\cite{Rossler_Nature2006,Muhlbauer_Science2009,Yu_Nature2010,Mohanta_SciRep2017} that can also be utilized in the proposed device geometry, instead of the combination of the ferromagnet and the heavy compound. Also, the SkX can be artificially-created without the need for any external magnetic field by nanopatternization~\cite{Sun_PRL2013}.

\begin{figure*}[t]
\begin{center}
\epsfig{file=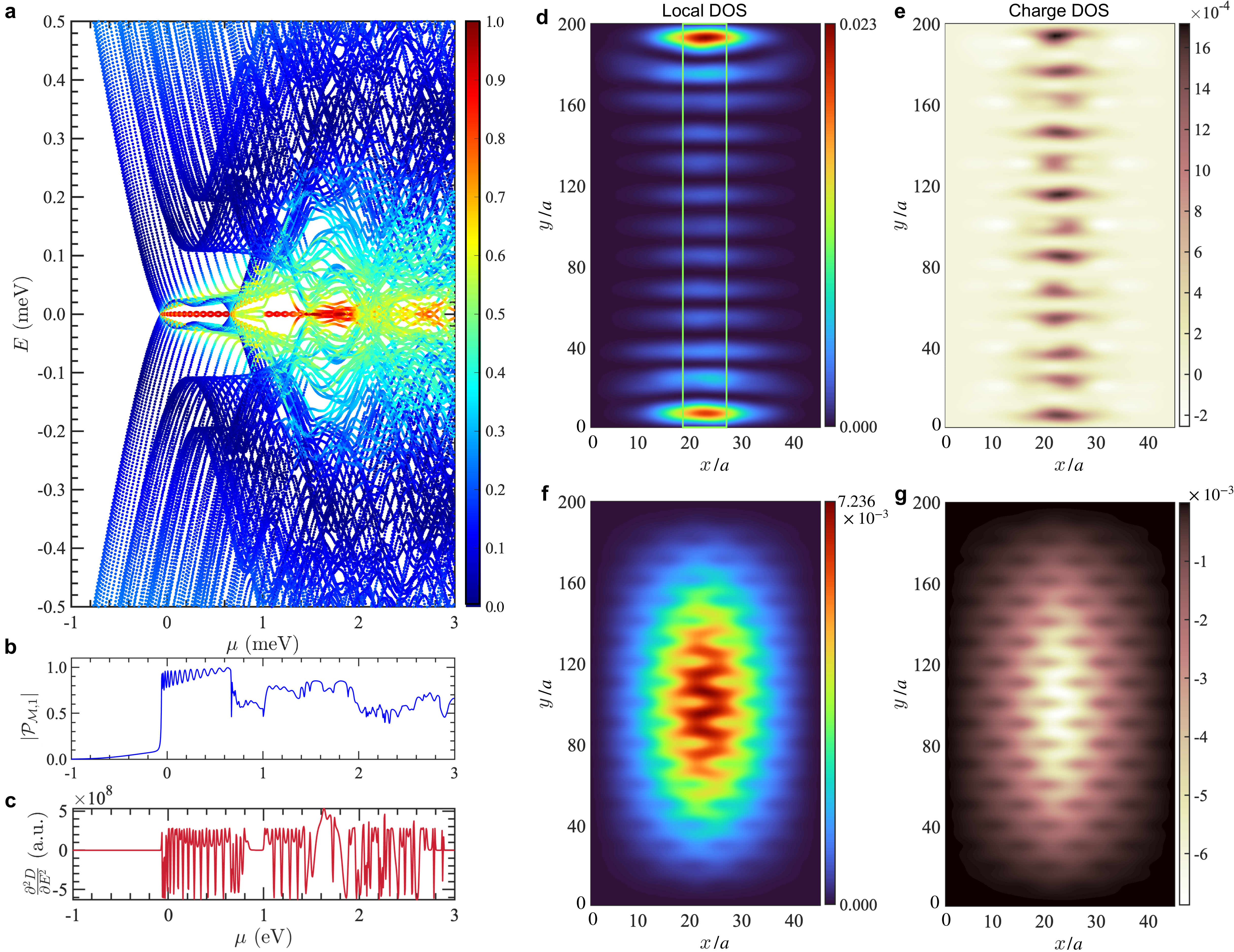,trim=0.0in 0.0in 0.0in 0.0in,clip=false, width=180mm}
\caption{{\bf Emergence of the Majorana bound states with changing chemical potential.} {\bf a} The quasiparticle spectrum of the planar Josephson junction at  phase difference $\varphi \!=\!0$ with varying chemical potential ($\mu$), showing the emergence of the zero-energy Majorana bound states. The colorbar represents the Majorana polarization $|{\cal P}_{{\cal M},n}|$ that displays the Majorana character of the quasiparticle states. The skyrmion crystal with a skyrmion diameter $D_{sk}=10a$ was obtained using a magnetic field $H_z\!=\!0.95J$ and a Dzyaloshinskii-Moriya interaction strength $D\!=\!1.6J$ in the Monte Carlo calculations. {\bf b}, {\bf c} The Majorana polarization $|{\cal P}_{{\cal M},1}|$ of the first positive eigenstate and the curvature of the density of states at zero energy $\frac{\partial^2 D}{\partial E^2}$, with varying $\mu$, showing the $\mu$ range within which the Majorana bound states appear. The delta-function-like peaks are associated with the oscillations of the MBS with changing $\mu$. {\bf d}, {\bf e} The profiles of the local density of states (DOS) and the charge DOS at $\mu \!=\!0.5$~meV. The green rectangle in {\bf d} indicates the quasi-one-dimensional metallic channel of the planar Josephson junction at the ends of which the Majorana bound states appear. {\bf f}, {\bf g} The profiles of the local DOS and the charge DOS at $\mu \!=\!-0.5$~meV, in the non-topological regime.}
\label{fig2}
\vspace{-7mm}
\end{center}
\end{figure*}
The spin texture $\mathbf{B}_i$ on the top layer of the ferromagnet, obtained from the Monte Carlo simulations, is used to obtain the low-energy spectrum of the planar Josephson junction by solving self-consistently the Bogoliubov-de Gennes equations. Since the SkX lies underneath the two-dimensional electron gas without a finite separation between them, it is reasonable to assume that the deviation in the magnetic fringing field in the two-dimensional electron gas from the original SkX texture is negligible. The proximity-induced superconductivity in the two-dimensional electron gas, which is subject to the SkX spin texture $\mathbf{B}_i$, is described by the Hamiltonian
\begin{align}
&\mathcal{H}_{\rm BdG}\!=\!-t\sum_{\langle ij \rangle,\sigma}(c_{i\sigma}^{\dagger}c_{j\sigma}+H.c.)+\sum_{i,\sigma}(4t-\mu)c_{i\sigma}^{\dagger}c_{i\sigma} \nonumber \\
&-\frac{1}{2}g\mu_{B}\sum_{i,\sigma}(\mathbf{B}_i \cdot \boldsymbol{\sigma})_{\sigma \sigma^{\prime}}c_{i\sigma}^{\dagger}c_{i\sigma^{\prime}}
+\sum_{i}(\Delta_i c_{i\uparrow}^{\dagger}c_{i\downarrow}^{\dagger}  +H.c.),
\label{H_bdg}
\end{align}
where $t=\hbar^2/(2m^{*}a^2)$ is the hopping energy, $m^*$ is the effective mass of electrons, $a$ is the unit spacing of the lattice grid, $\mu$ is the chemical potential, and $\Delta_i$ is the induced local $s$-wave pairing amplitude on the two sides of the Josephson junction that are attached to the top Al layer. The pairing amplitude $\Delta_i \!=\!-U_i\langle c_{i\uparrow}c_{i\downarrow} \rangle$ is calculated self-consistently using the onsite attractive interaction strength $U_i$ of the induced superconducting states in the two-dimensional electron gas. $U_i=U$ in the two-dimensional electron gas below the Al superconductors and zero in the middle metallic channel. The value $U\!=\!2$~meV is determined by setting $\Delta_i \!=\!0.2$~meV, the estimated proximity-induced gap magnitude for a two-dimensional electron gas with an SC interface~\cite{Gul_NanoLett2017,Deng_Science2016}, without any spin texture. The $g$ factor and the effective mass are set to $g\!=\!50$ and $m^*\!=\!0.017m_0$ for InSb~\cite{Nedniyom_PRB2009,Qu_NanoLett2016}. The lattice grid spacing used is $a\!=\!10$ nm~\cite{gridspacing} with which the hopping energy becomes $t\!=\!22.44$~meV. The amplitude of the spin texture $\mathbf{B}_i$ is set to $B_0\!=\!0.3$~T, compatible with the saturation magnetization $M_s\!=\!1.7\times10^6$~A/m for CoFe~\cite{Mohanta_PRApplied2019,Tong_PRB2019}. HM/ferromagnet interfaces with Pt, Pd, Ag, Ir, and Au as the HM have been developed together with Co, Fe, and their alloys (see \textit{e.g.} Ref.~\onlinecite{Zeng_APL2019}). We present results for a planar Josephson junction  with length $L_y\!=\!2$~$\mu$m, transverse length of the SC leads $L_x\!=\!200$~nm, and width of the quasi-one-dimensional metallic channel $W\!=\!50$~nm.\\

\begin{figure*}[t]
\begin{center}
\epsfig{file=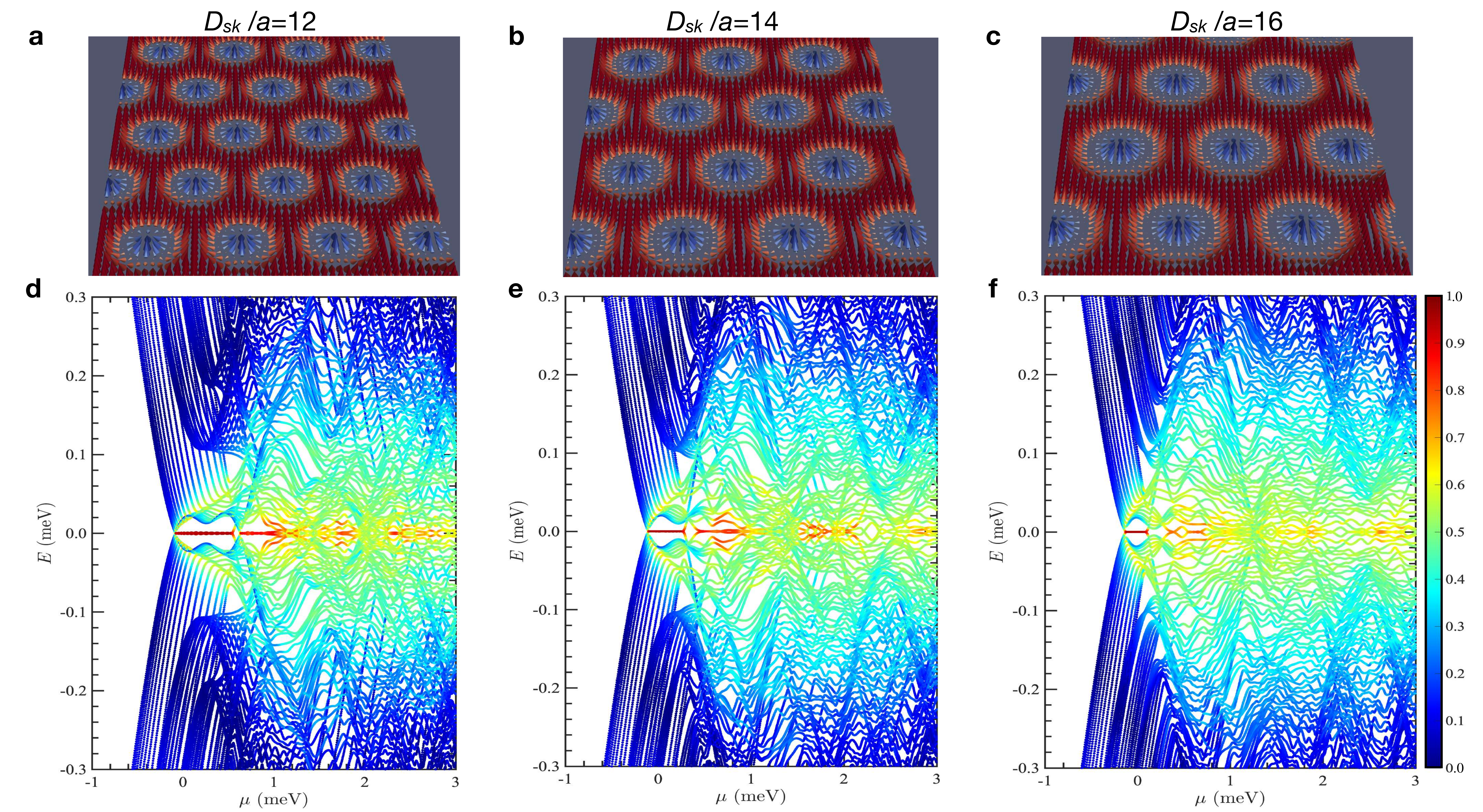,trim=0.0in 0.0in 0.0in 0.0in,clip=false, width=180mm}
\caption{{\bf Skyrmion control of the Majorana bound states.} {\bf a}-{\bf c} The skyrmion crystals of different skyrmion diameters {\bf a} $D_{sk}\!=\!12a$, {\bf b} $D_{sk}\!=\!14a$, {\bf c} $D_{sk}\!=\!16a$, obtained in the Monte Carlo calculations using {\bf a} a magnetic field $H_z\!=\!0.8J$, a Dzyaloshinskii-Moriya interaction strength $D\!=\!1.4J$, {\bf b} $H_z\!=\!0.45J$, $D\!=\!J$, and {\bf c} $H_z\!=\!0.23J$, $D\!=\!0.6J$.  {\bf d}-{\bf f} The corresponding quasiparticle spectra of the planar Josephson junction at the phase difference $\varphi \!=\!0$ with varying chemical potential, obtained by solving the Bogoliubov-de Gennes equations with the above skyrmion crystal spin configurations. The colorbar in  {\bf d}-{\bf f} represents the Majorana polarization $|{\cal P}_{{\cal M},n}|$ of the quasiparticle states.}
\label{fig3}
\vspace{-7mm}
\end{center}
\end{figure*}

\noindent {\bf Emergence of the MBS.} The low-energy spectrum, shown in Fig.~\ref{fig2}{\bf a}, reveals that there exist multiple ranges of the chemical potential within which the zero-energy MBS appear. To determine the Majorana character of the quasiparticle states, we compute the Majorana polarization, defined as~\cite{Sticlet_PRL2012,Sedlmayr_PRB2015}
\begin{align}
{\cal P_M}_{,n}=2\sum_{i} u_{i\downarrow}^{n}v_{i\downarrow}^{n*}-u_{i\uparrow}^{n}v_{i\uparrow}^{n*},
\label{Majpol}
\end{align}
where $u_{i\uparrow}^{n}$ and $v_{i\uparrow}^{n}$ are the Bogoliubov-de Gennes quasiparticle and quasihole amplitudes, respectively, corresponding to the $n^{\rm th}$ eigenstate, spin $\uparrow$, and site $i$.  As evident from Fig.~\ref{fig2}{\bf a}, ${\cal |P_M}_{,n}|\approx1$ indicates the occurrence of a pair of robust MBS with a finite topological energy gap. The Majorana polarization $|{\cal P}_{{\cal M},1}|$ of the first positive eigenstate, plotted with $\mu$ in Fig.~\ref{fig2}{\bf b}, acquires finite values within the range of $\mu$, in which the MBS emerge. The delta function-like peaks in $|{\cal P}_{{\cal M},1}|$ are the signatures of the Majorana oscillations, which is also clearly seen in the low-energy spectrum in Fig.~\ref{fig2}{\bf a}, originating due to the overlap of the MBS wave functions at the two ends of the finite-length quasi-one-dimensional channel. The Majorana oscillations in $|{\cal P}_{{\cal M},1}|$ have also been confirmed from the calculations of a one-dimensional wire (for results in the wire geometry, see Supplementary Note 4). The Majorana polarization, with a modification in the expression used in Eq.~\ref{Majpol}, was proposed to be probed in this planar Josephson junction geometry using the spin-selective Andreev reflection technique~\cite{Godzik_PRB2020}. To further characterize the evolution of the topological superconductivity with changing a parameter, such as $\mu$, we look at the curvature of the density of states at zero energy $\frac{\partial^2 D}{\partial E^2}$, where $D(E)$ is defined as~\cite{Scharf_PRB2019}
\begin{align}
D(E)=\sum_{i,n,\sigma} (|u_{i\sigma}^{n}|^2+|v_{i\sigma}^{n}|^2) \delta(E-E_n),
\end{align}
and $\delta(E-E_n)$ is modeled using a Gaussian with broadening $0.001$~meV ($\ll t$). The second derivative is computed using the second-order finite-difference method. These two quantities, $|{\cal P}_{{\cal M},1}|$ and $\frac{\partial^2 D}{\partial E^2}$, may provide additional insight in the experimental detection of the MBS, besides  the conventional zero-bias conductance peak~\cite{Sengupta_PRB2001} which often leads to ambiguity due to other possible zero-bias states in a superconductor~\cite{Kuerten_PRB2017}.

As shown in Fig.~\ref{fig2}{\bf c}, $\frac{\partial^2 D}{\partial E^2}$ takes finite values in the same ranges of $\mu$ as that of the Majorana polarization $|{\cal P}_{{\cal M},1}|$.  The Majorana oscillations, in the form of delta-function-like peaks, is also noticeable in $\frac{\partial^2 D}{\partial E^2}$, albeit with changes in the sign. To visualize the location of the zero-energy MBS, we show, in Fig.~\ref{fig2}{\bf d}, the profile of the local density of states $\rho_{_{\rm LDOS}}^i\!=\!\sum_{\sigma}(|u_{i\sigma}|^2+|v_{i\sigma}|^2)$, corresponding to the lowest positive-energy eigenstate at $\mu\!=\!0.5$~meV where the Josephson junction is in the topological superconducting regime. The sharp peaks in $\rho_{_{\rm LDOS}}^i$ indicate that the MBS are localized predominantly near the two ends of the quasi-one-dimensional channel. Figure~\ref{fig2}{\bf e} shows the profile of the charge density of states $\rho_{_{\rm CDOS}}^i\!=\!\sum_{\sigma}(|u_{i\sigma}|^2-|v_{i\sigma}|^2)$ corresponding to the lowest positive-energy eigenstate at $\mu\!=\!0.5$~meV. The profiles of $\rho_{_{\rm LDOS}}^i$ and $\rho_{_{\rm CDOS}}^i$, at $\mu\!=\!-0.5$~meV where the Josephson junction is in the topologically-trivial superconducting regime, are shown in Fig.~\ref{fig2}{\bf f} and Fig.~\ref{fig2}{\bf g}, respectively. In this case, both the quasiparticle state and the charge density are distributed near the middle of the quasi-one-dimensional channel. Interestingly, a comparison of Fig.~\ref{fig2}{\bf e} and Fig.~\ref{fig2}{\bf g}, implies an order-of-magnitude suppression in $\rho_{_{\rm CDOS}}^i$ which is reminiscent of the local charge-neutrality signature of the MBS and is another confirmation of the Majorana character of this state. The above results establish that the spin-orbit coupling, generated by the SkX, alone can lead to the emergence of the MBS in the planar Josephson junction devices.\\

\noindent {\bf Skyrmion control.} The skyrmion size in a SkX is tunable, with remarkable precession, by an external magnetic field, magnetic anisotropy and advanced symmetry protocol at heterointerfaces~\cite{Soumyanarayanan_NMat2017,Skoropata_SciAdv3902,LoConte_NanoLett2020}. In our Monte Carlo simulations, the skyrmion size was varied by tuning the magnetic field and the DMI, as shown in Figs.~\ref{fig3}{\bf a}-{\bf c}. The Bogoliubov-de Gennes quasiparticle spectra at different skyrmion sizes, shown in Figs.~\ref{fig3}{\bf d}-{\bf f}, imply  that the presence of the zero energy MBS at a given chemical potential can be turned ON or OFF by only changing the skyrmion properties. The minimum diameter of the skyrmions, realized in our Monte Carlo simulations, is 10 lattice spacings for which the MBS appear in the discussed planar Josephson junction setting. With increasing the skyrmion size, we find that the  range of the chemical potential within which the MBS appear decreases effectively, however, the oscillation amplitude of the MBS is suppressed gradually. Therefore, the skyrmions offer a unique ability to manipulate the localization length of the MBS in the planar Josephson junctions. The strongly-localized MBS in the skyrmion-tuned planar Josephson junctions can, therefore, have advantageous over other platforms for MBS realization in fault-tolerant topological quantum computing. 

The broken inversion symmetry at the interface between the two-dimensional electron gas and the superconductor, often leads to a sizable intrinsic Rashba spin-orbit coupling which is usually considered as the primary mechanism for modifying the pairing symmetry of the induced superconductivity~\cite{Mohanta_EPL2014,Mohanta_PRB2018}, leading to the desired topological superconductivity. We find that the MBS remain robust in the presence of the intrinsic Rashba spin-orbit coupling (for details on the effect of Rashba spin-orbit coupling, see Supplementary Note 2). By placing the SkX texture only below the middle quasi-1D channel, we didn't find robust MBS formation (for details, see Supplementary Note 3) and hence it suggests the significance of placing the SkX texture below the entire Josephson junction.\\

\begin{figure}[t]
\begin{center}
\epsfig{file=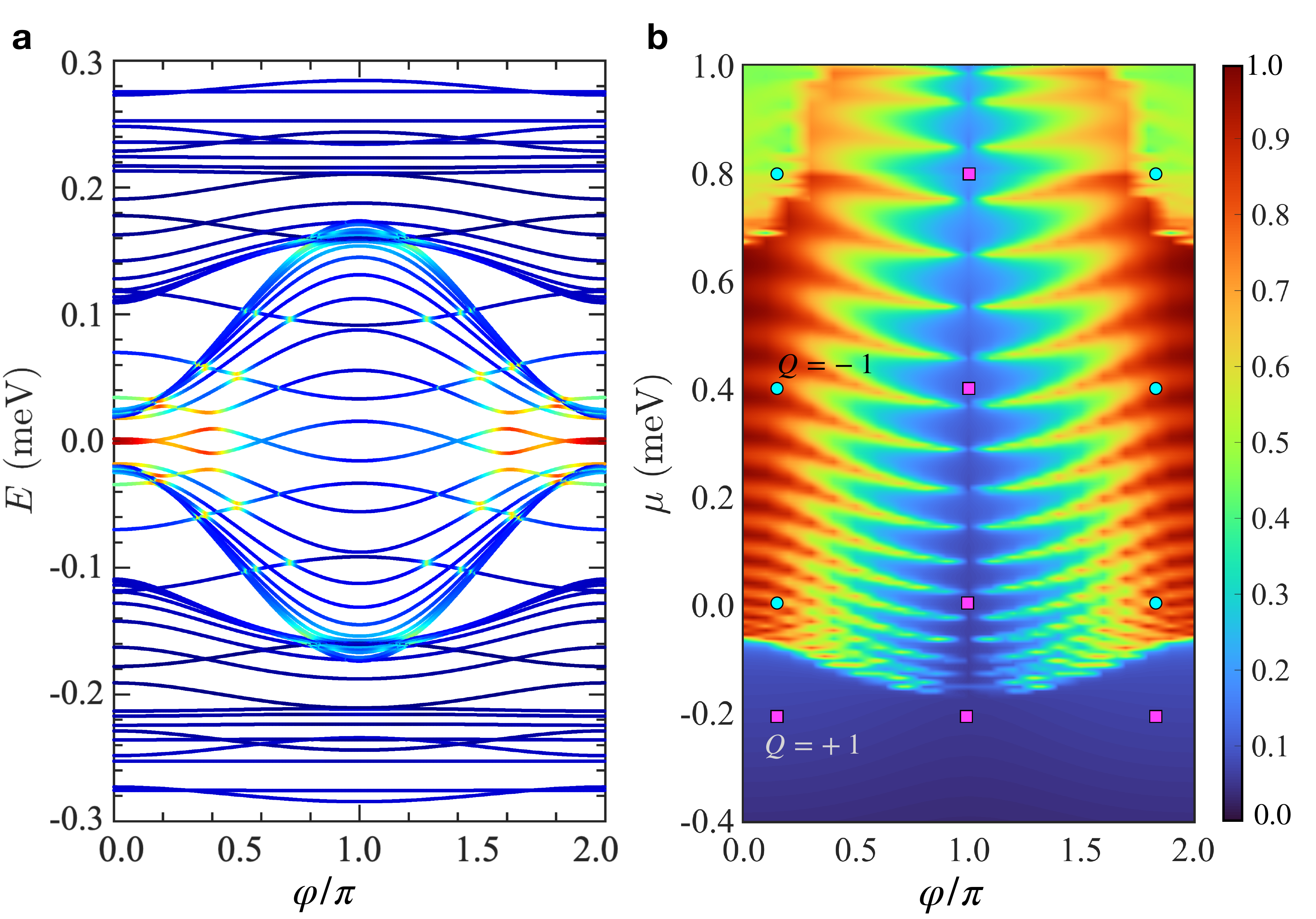,trim=0.0in 0.0in 0.0in 0.0in,clip=false, width=90mm}
\caption{{\bf Phase control of the Majorana bound states.} {\bf a} Quasiparticle spectrum of the planar Josephson junction at a chemical potential $\mu=0.5$~meV with changing phase difference $\varphi$ between the two superconducting regions. The skyrmion diameter of the skyrmion crystal was $D_{sk}\!=\!10a$. {\bf b} Colormap of the Majorana polarization of the first positive energy state in the plane spanned by the phase difference $\varphi$ and the chemical potential $\mu$. The Majorana bound states appear in the parameter regimes with $|{\cal P}_{{\cal M},1}| \approx$$1$, $\varphi=0$ being the most favorable scenario in this set up. The cyan and magenta dots in {\bf b} represent the topological invariant, respectively, $Q=-1$ (topologically nontrivial) and $Q=+1$ (topologically trivial).}
\label{fig4}
\vspace{-8mm}
\end{center}
\end{figure}
\noindent {\bf Phase control.} Another important control parameter, that sets the planar Josephson junctions apart from other platforms hosting the MBS, is the phase difference $\varphi$ between the two superconducting regions of a Josephson junction. The theoretical prediction~\cite{Pientka_PRX2017} and the subsequent experimental discoveries~\cite{Ren_Nature2019,Fornieri_Nature2019} suggest that the Josephson junction needs to be biased by a phase difference $\varphi=\pi$ to minimize the critical Zeeman field, required for inducing the topological superconductivity. Remarkably, in the current Josephson junction set up with the SkX, the topological superconductivity is induced at $\varphi=0$, as depicted by the quasiparticle spectrum with varying $\varphi$ in Fig.~\ref{fig4}{\bf a}. With increasing $\varphi$, the MBS move gradually from zero to higher energies, indicating an enhancement in the localization length of the MBS. The MBS appear again at zero energy above $\varphi \approx 3\pi/2$. 

This dephasing effect of the MBS can be understood from the Majorana oscillations -- as we find that the oscillation increases with increasing $\varphi$ in the range $0 < \varphi \leq \pi$ (for details, see Supplementary Note 1). The finite length of the quasi-one-dimensional metallic channel gives rise to the oscillations of the zero-energy MBS with varying chemical potential $\mu$. Furthermore, the finite width of the metallic channel provides extra room for delocalization of the MBS at the two ends, contributing additively to the Majorana oscillation. In the previous works on this geometry, a magnetic field is applied in the plane of the planar Josephson junction which locks the phases of the MBS to $\varphi=\pi$. In our set up, there is no magnetic field applied to a particular direction, but a chiral magnetism exists throughout the junction and, therefore, a $\pi$ phase biasing is not required in this case. The middle metallic region of the Josephson junction can be perceived as a quasi-one-dimensional void region surrounded by the superconducting two-dimensional electron gas. Additional phase difference between the two superconducting sides, therefore, only causes disruption to the induced topological superconductivity. This phenomenon generically takes place at several values of the chemical potential, as shown in Fig.~\ref{fig4}{\bf b}, where we plot the Majorana polarization $|{\cal P}_{{\cal M},1}|$ of the lowest positive energy eigenstate in the parameter space spanned by the phase difference $\varphi$ and the chemical potential $\mu$. For the chosen range of $\mu$ values, the Majorana polarization decreases substantially within the range $\pi/2 \lessapprox \varphi \lessapprox 3\pi/2$. The Majorana oscillation in $|{\cal P}_{{\cal M},1}|$, however, survives up to $\varphi \approx \pi$. At $\varphi=\pi$, $|{\cal P}_{{\cal M},1}|$ vanishes completely, indicating the disappearance of the MBS. Therefore, $\varphi=0$ is the most favorable condition to realize the MBS in our Josephson junction set up and the phase difference can be further tuned to control the presence of the MBS. To check the consistency of our assignment of MBS, we also compute the $Z_2$ topological index $Q$ at several points of the above phase diagram using an effective one-dimensional Hamiltonian of the planar Josephson junction with the SkX. In Fig.~\ref{fig4}{\bf b}, we show the topological invariant which confirms the phase diagram.

\noindent \textbf{\\Conclusion\\}
The skyrmions bring outstanding control functionalities to the planar Josephson junctions for the creation and manipulation of the zero-energy MBS and their localization properties. The SkXs, being realized in an abundance of magnetic materials and also artificially created in patterned magnetic materials, offer a feasible approach for advanced manipulation of the zero-energy MBS. The proposed planar Josephson junction, combined with a SkX, has the major advantages that there is no need for a strong intrinsic Rashba-type spin-orbit coupling and phase-biasing constraint for the realization of the zero-energy MBS. The enhanced tunability of the MBS in the proposed two-dimensional platform opens up opportunities for designing feasible MBS braiding protocols for the fault-tolerant topological quantum computing, and investigating Majorana spectroscopy using the multi-terminal superconducting quantum interference devices.\\
\noindent \small{\bf \\Methods}\\
\noindent {\bf Monte Carlo simulations.} The SkX spin configurations were obtained using a $L_x\times L_y \times L_z$ lattice with periodic boundary conditions along the $x$ and $y$ directions and open boundary conditions along the $z$ direction. A bias-free sampling method, that provides a full and uniform coverage of the phase space spanned by the spin angles, was used for generating the completely random spin configurations. The calculation was started at a high temperature $T \!=\! 10J$ with a random spin configuration and the temperature was lowered slowly down to a low value $T \!=\! 0.001J$ in $2000$ steps. At each temperature step, $10^{10}$ Monte Carlo spin updates were performed. In each spin update step, a new spin direction was chosen randomly within a small cone spanned around the initial spin direction. The new spin configuration was accepted or rejected according to the Metropolis energy-minimization algorithm by comparing the total energies of the previous and the new trial spin configurations, calculated using the Hamiltonian~(\ref{Ham}).\\
\noindent {\bf \\Self-consistent Bogoliubov-de Gennes formalism.} The Hamiltonian~(\ref{H_bdg}), which is quadratic in the fermionic operators $\hat{c}_{i\sigma}$, can be solved by exact diagonalization via a unitary transformation $\hat{c}_{i\sigma}=\sum_{n} u_{i\sigma}^{n} \hat{\gamma}_{n} +v_{i\sigma}^{n*} \hat{\gamma}_{n}^{\dagger}$, where $\hat{\gamma}_n^{\dagger}$ ($\hat{\gamma}_n$) is a fermionic creation (annihilation) operator of the quasiparticle/quasiphole state in the $n^{\rm th}$ energy eigenstate. The quasiparticle amplitudes $u_{i\sigma}^{n}$ and the quasihole amplitudes $u_{i\sigma}^{n}$ are determined by solving the Bogoliubov-de Gennes equations: $\sum_{j}\mathcal{H}_{ij} \psi_j^n=\epsilon_n\psi_n^i$, where $\psi_i^n=[u_{i\uparrow}^{n}, u_{i\downarrow}^{n}, v_{i\uparrow}^{n}, v_{i\downarrow}^{n}]^T$ is the basis wave function and $\epsilon_n$ is the $n^{\rm th}$ energy eigenvalue. The Hamiltonian $\mathcal{H}_{ij}$ is expressed in the following matrix form
\begin{align}
\mathcal{H}_{ij}=
\begin{pmatrix} \mathcal{H}_{\uparrow \uparrow} & \mathcal{H}_{\uparrow \downarrow} & 0 & \Delta_i \\  
 \mathcal{H}_{\downarrow \uparrow} & \mathcal{H}_{\downarrow \downarrow} & -\Delta_i & 0\\ 
  0 & -\Delta_i^* & -\mathcal{H}_{\uparrow \uparrow}^* & -\mathcal{H}_{\uparrow \downarrow}^* \\ 
   \Delta_i^* & 0 & -\mathcal{H}_{\downarrow \uparrow}^* & -\mathcal{H}_{\downarrow \downarrow}^* \\ 
   \end{pmatrix},
\end{align}
where $\mathcal{H}_{\uparrow \uparrow,\downarrow \downarrow}=-t(1-\delta_{ij})+(4t-\mu)\delta_{ij}-1/2\sigma g\mu_B B_z$, where $\sigma=\pm$ for $\uparrow \uparrow$, $\downarrow \downarrow$, and $\mathcal{H}_{\uparrow \downarrow}=-1/2 g\mu_B (B_x +iB_y)$.
The $s$-wave pairing amplitude $\Delta_{i}=-U_i\langle c_{i\uparrow}c_{i\downarrow} \rangle$ is computed using $\Delta_{i}=-U_i\sum_{n}\big[u_{i\uparrow}^n v_{i\downarrow}^{n*}(1-f(\epsilon_n)) + u_{i\downarrow}^n v_{i\uparrow}^{n*}f(\epsilon_n)\big] $,
where $U_i=U$ inside the two superconducting regions and zero in the middle quasi-one-dimensional metallic channel, $f(\epsilon_n)=1/(1+e^{\epsilon_n/k_{B}T})$ is the Fermi-Dirac distribution function at temperature $T$. The self-consistency iterations were performed until the pairing amplitudes $\Delta_{i}$ were converged at every lattice sites.\\
\noindent {\bf \\Calculation of topological invariant.} A triangular SkX can be described by the following spin structure
\begin{align}
\mathbf{S}_i=S(\sin{\theta_i} \cos{\phi_i}, ~\sin{\theta_i} \sin{\phi_i}, ~\cos{\theta_i}),
\end{align}
where $S$ is the spin amplitude and the spin angles $\theta_i$ and $\phi_i$ are defined as
\begin{align}
\theta_i=\pi ~{\rm min}\Big( \frac{|\mathbf{r}-\mathbf{R}_i|}{R},1 \Big),
\phi_i={\rm arctan}\Big( \frac{y-y_i}{x-x_i} \Big),
\end{align}
$\mathbf{R}_i=(x_i,y_i)$ is the skyrmion center near position $\mathbf{r}=(x,y)$ and $R$ is the  skyrmion radius.
To obtain an effective Hamiltonian for the planar JJ, we perform the following unitary transformation that rotates the local spin $\mathbf{S}_i$ with respect to the SkX plane normal, the $z$ direction
\begin{align}
\begin{pmatrix}
c_{i\uparrow}\\
c_{i\downarrow}
\end{pmatrix}
=\hat{\Gamma}_i
\begin{pmatrix}
d_{i\uparrow}\\
d_{i\downarrow}
\end{pmatrix}
\end{align}
where
\begin{align}
\hat{\Gamma}_i=\exp \Big ( -i \frac{\theta_i}{2} \frac{\hat{z}\times \mathbf{S}_i}{|\hat{z}\times \mathbf{S}_i|} \cdot \sigma \Big ).
\end{align}
The Hamiltonian (2) in the main text, in the transformed basis, becomes
\begin{align}
\mathcal{H}\!=&\!-\sum_{\langle ij \rangle,\sigma, \sigma^{\prime}} t^{\prime}_{\sigma \sigma^{\prime}}d_{i\sigma}^{\dagger}d_{j\sigma^{\prime}}+\sum_{i,\alpha}(4t-\mu-\frac{1}{2}g\mu_B B_0\sigma^z_{\sigma \sigma})d_{i\sigma}^{\dagger}d_{i\sigma} \nonumber \\
&+\sum_{i}(\Delta_i d_{i\uparrow}^{\dagger}d_{i\downarrow}^{\dagger}  +{\rm H.c.}),
\end{align}
where the new hopping integral is given by
\begin{align}
t^{\prime}=
\begin{pmatrix}
-t & -\alpha^*\\
\alpha & -t^*
\end{pmatrix},
\end{align}
$\alpha$ being the strength of the generated SOC. 

To obtain an effective one-dimensional Hamiltonian, we consider an infinitely long junction (\textit{i.e.} $L_y\rightarrow \infty$) and perform a partial Fourier transform $d_{i,k_y,\sigma}=\sum_{j}e^{ik_y y}d_{i,j,\sigma}$. The resulting Hamiltonian $\mathcal{H}(x,k_y)$ is then used to compute the topological invariant $Q$, the $Z_2$ topological index associated with the broken chiral symmetry, given by~\cite{Tewari_PRL2012}
\begin{align}
Q={\rm sgn } \Big[ \frac{{\rm Pf}\{\mathcal{H}(x,k_y=\pi)\}}{{\rm Pf}\{\mathcal{H}(x,k_y=0)\}} \Big]
\end{align}
where  `Pf' denotes the Pfaffian. A topologically nontrivial phase is determined by $Q=-1$, while $Q=1$ represents the trivial phase.

\noindent  \textbf{\\Acknowledgements}\\
This work was supported by the U.S. Department of Energy, Office of Science, Basic Energy Sciences, Materials Sciences and Engineering Division. \\



\vspace{-4mm}
\def\bibsection{\section*{\refname}} 

\begin{thebibliography}{58}%
\makeatletter
\providecommand \@ifxundefined [1]{%
 \@ifx{#1\undefined}
}%
\providecommand \@ifnum [1]{%
 \ifnum #1\expandafter \@firstoftwo
 \else \expandafter \@secondoftwo
 \fi
}%
\providecommand \@ifx [1]{%
 \ifx #1\expandafter \@firstoftwo
 \else \expandafter \@secondoftwo
 \fi
}%
\providecommand \natexlab [1]{#1}%
\providecommand \enquote  [1]{``#1''}%
\providecommand \bibnamefont  [1]{#1}%
\providecommand \bibfnamefont [1]{#1}%
\providecommand \citenamefont [1]{#1}%
\providecommand \href@noop [0]{\@secondoftwo}%
\providecommand \href [0]{\begingroup \@sanitize@url \@href}%
\providecommand \@href[1]{\@@startlink{#1}\@@href}%
\providecommand \@@href[1]{\endgroup#1\@@endlink}%
\providecommand \@sanitize@url [0]{\catcode `\\12\catcode `\$12\catcode
  `\&12\catcode `\#12\catcode `\^12\catcode `\_12\catcode `\%12\relax}%
\providecommand \@@startlink[1]{}%
\providecommand \@@endlink[0]{}%
\providecommand \url  [0]{\begingroup\@sanitize@url \@url }%
\providecommand \@url [1]{\endgroup\@href {#1}{\urlprefix }}%
\providecommand \urlprefix  [0]{URL }%
\providecommand \Eprint [0]{\href }%
\providecommand \doibase [0]{http://dx.doi.org/}%
\providecommand \selectlanguage [0]{\@gobble}%
\providecommand \bibinfo  [0]{\@secondoftwo}%
\providecommand \bibfield  [0]{\@secondoftwo}%
\providecommand \translation [1]{[#1]}%
\providecommand \BibitemOpen [0]{}%
\providecommand \bibitemStop [0]{}%
\providecommand \bibitemNoStop [0]{.\EOS\space}%
\providecommand \EOS [0]{\spacefactor3000\relax}%
\providecommand \BibitemShut  [1]{\csname bibitem#1\endcsname}%
\let\auto@bib@innerbib\@empty
\bibitem [{\citenamefont {Kitaev}(2003)}]{Kitaev_AnnPhys2003}%
  \BibitemOpen
  \bibfield  {author} {\bibinfo {author} {\bibfnamefont {A.Yu.}\ \bibnamefont
  {Kitaev}},\ }\bibfield  {title} {\enquote {\bibinfo {title} {{Fault-tolerant
  quantum computation by anyons}},}\ }\href
  {http://www.sciencedirect.com/science/article/pii/S0003491602000180}
  {\bibfield  {journal} {\bibinfo  {journal} {Ann. Phys.}\ }\textbf {\bibinfo
  {volume} {303}},\ \bibinfo {pages} {2} (\bibinfo {year} {2003})}\BibitemShut
  {NoStop}%
\bibitem [{\citenamefont {Desjardins}\ \emph {et~al.}(2019)\citenamefont
  {Desjardins}, \citenamefont {Contamin}, \citenamefont {Delbecq},
  \citenamefont {Dartiailh}, \citenamefont {Bruhat}, \citenamefont {Cubaynes},
  \citenamefont {Viennot}, \citenamefont {Mallet}, \citenamefont {Rohart},
  \citenamefont {Thiaville}, \citenamefont {Cottet},\ and\ \citenamefont
  {Kontos}}]{Desjardins_NatMat2019}%
  \BibitemOpen
  \bibfield  {author} {\bibinfo {author} {\bibfnamefont {M.~M.}\ \bibnamefont
  {Desjardins}}, \bibinfo {author} {\bibfnamefont {L.~C.}\ \bibnamefont
  {Contamin}}, \bibinfo {author} {\bibfnamefont {M.~R.}\ \bibnamefont
  {Delbecq}}, \bibinfo {author} {\bibfnamefont {M.~C.}\ \bibnamefont
  {Dartiailh}}, \bibinfo {author} {\bibfnamefont {L.~E.}\ \bibnamefont
  {Bruhat}}, \bibinfo {author} {\bibfnamefont {T.}~\bibnamefont {Cubaynes}},
  \bibinfo {author} {\bibfnamefont {J.~J.}\ \bibnamefont {Viennot}}, \bibinfo
  {author} {\bibfnamefont {F.}~\bibnamefont {Mallet}}, \bibinfo {author}
  {\bibfnamefont {S.}~\bibnamefont {Rohart}}, \bibinfo {author} {\bibfnamefont
  {A.}~\bibnamefont {Thiaville}}, \bibinfo {author} {\bibfnamefont
  {A.}~\bibnamefont {Cottet}}, \ and\ \bibinfo {author} {\bibfnamefont
  {T.}~\bibnamefont {Kontos}},\ }\bibfield  {title} {\enquote {\bibinfo {title}
  {Synthetic spin--orbit interaction for {M}ajorana devices},}\ }\href
  {https://doi.org/10.1038/s41563-019-0457-6} {\bibfield  {journal} {\bibinfo
  {journal} {Nat. Mater.}\ }\textbf {\bibinfo {volume} {18}},\ \bibinfo {pages}
  {1060--1064} (\bibinfo {year} {2019})}\BibitemShut {NoStop}%
\bibitem [{\citenamefont {Yazdani}(2019)}]{Yazdani_NMat2019}%
  \BibitemOpen
  \bibfield  {author} {\bibinfo {author} {\bibfnamefont {A.}~\bibnamefont
  {Yazdani}},\ }\bibfield  {title} {\enquote {\bibinfo {title} {Conjuring
  {M}ajorana with synthetic magnetism},}\ }\href
  {https://doi.org/10.1038/s41563-019-0477-2} {\bibfield  {journal} {\bibinfo
  {journal} {Nat. Mater.}\ }\textbf {\bibinfo {volume} {18}},\ \bibinfo {pages}
  {1036} (\bibinfo {year} {2019})}\BibitemShut {NoStop}%
\bibitem [{\citenamefont {Nayak}\ \emph {et~al.}(2008)\citenamefont {Nayak},
  \citenamefont {Simon}, \citenamefont {Stern}, \citenamefont {Freedman},\ and\
  \citenamefont {Das~Sarma}}]{Nayak_RMP2008}%
  \BibitemOpen
  \bibfield  {author} {\bibinfo {author} {\bibfnamefont {C.}~\bibnamefont
  {Nayak}}, \bibinfo {author} {\bibfnamefont {S.~H.}\ \bibnamefont {Simon}},
  \bibinfo {author} {\bibfnamefont {A.}~\bibnamefont {Stern}}, \bibinfo
  {author} {\bibfnamefont {M.}~\bibnamefont {Freedman}}, \ and\ \bibinfo
  {author} {\bibfnamefont {S.}~\bibnamefont {Das~Sarma}},\ }\bibfield  {title}
  {\enquote {\bibinfo {title} {Non-abelian anyons and topological quantum
  computation},}\ }\href {https://link.aps.org/doi/10.1103/RevModPhys.80.1083}
  {\bibfield  {journal} {\bibinfo  {journal} {Rev. Mod. Phys.}\ }\textbf
  {\bibinfo {volume} {80}},\ \bibinfo {pages} {1083--1159} (\bibinfo {year}
  {2008})}\BibitemShut {NoStop}%
\bibitem [{\citenamefont {Alicea}\ \emph {et~al.}(2011)\citenamefont {Alicea},
  \citenamefont {Oreg}, \citenamefont {Refael}, \citenamefont {von Oppen},\
  and\ \citenamefont {Fisher}}]{Alicea_NatPhy2011}%
  \BibitemOpen
  \bibfield  {author} {\bibinfo {author} {\bibfnamefont {J.}~\bibnamefont
  {Alicea}}, \bibinfo {author} {\bibfnamefont {Y.}~\bibnamefont {Oreg}},
  \bibinfo {author} {\bibfnamefont {G.}~\bibnamefont {Refael}}, \bibinfo
  {author} {\bibfnamefont {F.}~\bibnamefont {von Oppen}}, \ and\ \bibinfo
  {author} {\bibfnamefont {M.~P.~A.}\ \bibnamefont {Fisher}},\ }\bibfield
  {title} {\enquote {\bibinfo {title} {{Non-Abelian statistics and topological
  quantum information processing in 1D wire networks}},}\ }\href
  {https://doi.org/10.1038/nphys1915} {\bibfield  {journal} {\bibinfo
  {journal} {Nat. Phys.}\ }\textbf {\bibinfo {volume} {7}},\ \bibinfo {pages}
  {412} (\bibinfo {year} {2011})}\BibitemShut {NoStop}%
\bibitem [{\citenamefont {Kjaergaard}\ \emph {et~al.}(2012)\citenamefont
  {Kjaergaard}, \citenamefont {W\"olms},\ and\ \citenamefont
  {Flensberg}}]{Kjaergaard_PRB2012}%
  \BibitemOpen
  \bibfield  {author} {\bibinfo {author} {\bibfnamefont {M.}~\bibnamefont
  {Kjaergaard}}, \bibinfo {author} {\bibfnamefont {K.}~\bibnamefont {W\"olms}},
  \ and\ \bibinfo {author} {\bibfnamefont {K.}~\bibnamefont {Flensberg}},\
  }\bibfield  {title} {\enquote {\bibinfo {title} {{Majorana fermions in
  superconducting nanowires without spin-orbit coupling}},}\ }\href
  {https://link.aps.org/doi/10.1103/PhysRevB.85.020503} {\bibfield  {journal}
  {\bibinfo  {journal} {Phys. Rev. B}\ }\textbf {\bibinfo {volume} {85}},\
  \bibinfo {pages} {020503} (\bibinfo {year} {2012})}\BibitemShut {NoStop}%
\bibitem [{\citenamefont {Klinovaja}\ \emph {et~al.}(2013)\citenamefont
  {Klinovaja}, \citenamefont {Stano}, \citenamefont {Yazdani},\ and\
  \citenamefont {Loss}}]{Klinovaja_PRL2013}%
  \BibitemOpen
  \bibfield  {author} {\bibinfo {author} {\bibfnamefont {J.}~\bibnamefont
  {Klinovaja}}, \bibinfo {author} {\bibfnamefont {P.}~\bibnamefont {Stano}},
  \bibinfo {author} {\bibfnamefont {A.}~\bibnamefont {Yazdani}}, \ and\
  \bibinfo {author} {\bibfnamefont {D.}~\bibnamefont {Loss}},\ }\bibfield
  {title} {\enquote {\bibinfo {title} {Topological superconductivity and
  {M}ajorana fermions in {RKKY} systems},}\ }\href
  {https://link.aps.org/doi/10.1103/PhysRevLett.111.186805} {\bibfield
  {journal} {\bibinfo  {journal} {Phys. Rev. Lett.}\ }\textbf {\bibinfo
  {volume} {111}},\ \bibinfo {pages} {186805} (\bibinfo {year}
  {2013})}\BibitemShut {NoStop}%
\bibitem [{\citenamefont {Klinovaja}\ and\ \citenamefont
  {Loss}(2013)}]{Klinovaja_PRX2013}%
  \BibitemOpen
  \bibfield  {author} {\bibinfo {author} {\bibfnamefont {J.}~\bibnamefont
  {Klinovaja}}\ and\ \bibinfo {author} {\bibfnamefont {D.}~\bibnamefont
  {Loss}},\ }\bibfield  {title} {\enquote {\bibinfo {title} {Giant spin-orbit
  interaction due to rotating magnetic fields in {G}raphene nanoribbons},}\
  }\href {https://link.aps.org/doi/10.1103/PhysRevX.3.011008} {\bibfield
  {journal} {\bibinfo  {journal} {Phys. Rev. X}\ }\textbf {\bibinfo {volume}
  {3}},\ \bibinfo {pages} {011008} (\bibinfo {year} {2013})}\BibitemShut
  {NoStop}%
\bibitem [{\citenamefont {Fatin}\ \emph {et~al.}(2016)\citenamefont {Fatin},
  \citenamefont {Matos-Abiague}, \citenamefont {Scharf},\ and\ \citenamefont
  {\ifmmode \check{Z}\else \v{Z}\fi{}uti\ifmmode~\acute{c}\else
  \'{c}\fi{}}}]{Geoffrey_PRL2016}%
  \BibitemOpen
  \bibfield  {author} {\bibinfo {author} {\bibfnamefont {G.~L.}\ \bibnamefont
  {Fatin}}, \bibinfo {author} {\bibfnamefont {A.}~\bibnamefont
  {Matos-Abiague}}, \bibinfo {author} {\bibfnamefont {B.}~\bibnamefont
  {Scharf}}, \ and\ \bibinfo {author} {\bibfnamefont {I.}~\bibnamefont
  {\ifmmode \check{Z}\else \v{Z}\fi{}uti\ifmmode~\acute{c}\else \'{c}\fi{}}},\
  }\bibfield  {title} {\enquote {\bibinfo {title} {Wireless {M}ajorana bound
  states: {F}rom magnetic tunability to braiding},}\ }\href
  {https://link.aps.org/doi/10.1103/PhysRevLett.117.077002} {\bibfield
  {journal} {\bibinfo  {journal} {Phys. Rev. Lett.}\ }\textbf {\bibinfo
  {volume} {117}},\ \bibinfo {pages} {077002} (\bibinfo {year}
  {2016})}\BibitemShut {NoStop}%
\bibitem [{\citenamefont {Mohanta}\ \emph
  {et~al.}(2019{\natexlab{a}})\citenamefont {Mohanta}, \citenamefont {Zhou},
  \citenamefont {Xu}, \citenamefont {Han}, \citenamefont {Kent}, \citenamefont
  {Shabani}, \citenamefont {\ifmmode \check{Z}\else
  \v{Z}\fi{}uti\ifmmode~\acute{c}\else \'{c}\fi{}},\ and\ \citenamefont
  {Matos-Abiague}}]{Mohanta_PRApplied2019}%
  \BibitemOpen
  \bibfield  {author} {\bibinfo {author} {\bibfnamefont {N.}~\bibnamefont
  {Mohanta}}, \bibinfo {author} {\bibfnamefont {T.}~\bibnamefont {Zhou}},
  \bibinfo {author} {\bibfnamefont {J.-W.}\ \bibnamefont {Xu}}, \bibinfo
  {author} {\bibfnamefont {J.~E.}\ \bibnamefont {Han}}, \bibinfo {author}
  {\bibfnamefont {A.~D.}\ \bibnamefont {Kent}}, \bibinfo {author}
  {\bibfnamefont {J.}~\bibnamefont {Shabani}}, \bibinfo {author} {\bibfnamefont
  {I.}~\bibnamefont {\ifmmode \check{Z}\else
  \v{Z}\fi{}uti\ifmmode~\acute{c}\else \'{c}\fi{}}}, \ and\ \bibinfo {author}
  {\bibfnamefont {Alex}\ \bibnamefont {Matos-Abiague}},\ }\bibfield  {title}
  {\enquote {\bibinfo {title} {{Electrical Control of Majorana Bound States
  Using Magnetic Stripes}},}\ }\href
  {https://link.aps.org/doi/10.1103/PhysRevApplied.12.034048} {\bibfield
  {journal} {\bibinfo  {journal} {Phys. Rev. Applied}\ }\textbf {\bibinfo
  {volume} {12}},\ \bibinfo {pages} {034048} (\bibinfo {year}
  {2019}{\natexlab{a}})}\BibitemShut {NoStop}%
\bibitem [{\citenamefont {Zhou}\ \emph {et~al.}(2019)\citenamefont {Zhou},
  \citenamefont {Mohanta}, \citenamefont {Han}, \citenamefont {Matos-Abiague},\
  and\ \citenamefont {\ifmmode \check{Z}\else
  \v{Z}\fi{}uti\ifmmode~\acute{c}\else \'{c}\fi{}}}]{Tong_PRB2019}%
  \BibitemOpen
  \bibfield  {author} {\bibinfo {author} {\bibfnamefont {T.}~\bibnamefont
  {Zhou}}, \bibinfo {author} {\bibfnamefont {N.}~\bibnamefont {Mohanta}},
  \bibinfo {author} {\bibfnamefont {J.~E.}\ \bibnamefont {Han}}, \bibinfo
  {author} {\bibfnamefont {A.}~\bibnamefont {Matos-Abiague}}, \ and\ \bibinfo
  {author} {\bibfnamefont {I.}~\bibnamefont {\ifmmode \check{Z}\else
  \v{Z}\fi{}uti\ifmmode~\acute{c}\else \'{c}\fi{}}},\ }\bibfield  {title}
  {\enquote {\bibinfo {title} {Tunable magnetic textures in spin valves: {F}rom
  spintronics to {M}ajorana bound states},}\ }\href
  {https://link.aps.org/doi/10.1103/PhysRevB.99.134505} {\bibfield  {journal}
  {\bibinfo  {journal} {Phys. Rev. B}\ }\textbf {\bibinfo {volume} {99}},\
  \bibinfo {pages} {134505} (\bibinfo {year} {2019})}\BibitemShut {NoStop}%
\bibitem [{\citenamefont {{Herbrych}}\ \emph {et~al.}(2020)\citenamefont
  {{Herbrych}}, \citenamefont {{{\'S}roda}}, \citenamefont {{Alvarez}},
  \citenamefont {{Mierzejewski}},\ and\ \citenamefont
  {{Dagotto}}}]{Elbio_arxiv2020}%
  \BibitemOpen
  \bibfield  {author} {\bibinfo {author} {\bibfnamefont {J.}~\bibnamefont
  {{Herbrych}}}, \bibinfo {author} {\bibfnamefont {M.}~\bibnamefont
  {{{\'S}roda}}}, \bibinfo {author} {\bibfnamefont {G.}~\bibnamefont
  {{Alvarez}}}, \bibinfo {author} {\bibfnamefont {M.}~\bibnamefont
  {{Mierzejewski}}}, \ and\ \bibinfo {author} {\bibfnamefont {E.}~\bibnamefont
  {{Dagotto}}},\ }\bibfield  {title} {\enquote {\bibinfo {title}
  {{Interaction-induced topological phase transition and Majorana edge states
  in low-dimensional orbital-selective Mott insulators}},}\ }\href
  {https://arxiv.org/abs/2011.05646} {\bibfield  {journal} {\bibinfo  {journal}
  {arXiv:2011.05646}\ } (\bibinfo {year} {2020})}\BibitemShut {NoStop}%
\bibitem [{\citenamefont {Shabani}\ \emph {et~al.}(2016)\citenamefont
  {Shabani}, \citenamefont {Kjaergaard}, \citenamefont {Suominen},
  \citenamefont {Kim}, \citenamefont {Nichele}, \citenamefont {Pakrouski},
  \citenamefont {Stankevic}, \citenamefont {Lutchyn}, \citenamefont
  {Krogstrup}, \citenamefont {Feidenhans'l}, \citenamefont {Kraemer},
  \citenamefont {Nayak}, \citenamefont {Troyer}, \citenamefont {Marcus},\ and\
  \citenamefont {Palmstr\o{}m}}]{Shabani_PRB2016}%
  \BibitemOpen
  \bibfield  {author} {\bibinfo {author} {\bibfnamefont {J.}~\bibnamefont
  {Shabani}}, \bibinfo {author} {\bibfnamefont {M.}~\bibnamefont {Kjaergaard}},
  \bibinfo {author} {\bibfnamefont {H.~J.}\ \bibnamefont {Suominen}}, \bibinfo
  {author} {\bibfnamefont {Younghyun}\ \bibnamefont {Kim}}, \bibinfo {author}
  {\bibfnamefont {F.}~\bibnamefont {Nichele}}, \bibinfo {author} {\bibfnamefont
  {K.}~\bibnamefont {Pakrouski}}, \bibinfo {author} {\bibfnamefont
  {T.}~\bibnamefont {Stankevic}}, \bibinfo {author} {\bibfnamefont {R.~M.}\
  \bibnamefont {Lutchyn}}, \bibinfo {author} {\bibfnamefont {P.}~\bibnamefont
  {Krogstrup}}, \bibinfo {author} {\bibfnamefont {R.}~\bibnamefont
  {Feidenhans'l}}, \bibinfo {author} {\bibfnamefont {S.}~\bibnamefont
  {Kraemer}}, \bibinfo {author} {\bibfnamefont {C.}~\bibnamefont {Nayak}},
  \bibinfo {author} {\bibfnamefont {M.}~\bibnamefont {Troyer}}, \bibinfo
  {author} {\bibfnamefont {C.~M.}\ \bibnamefont {Marcus}}, \ and\ \bibinfo
  {author} {\bibfnamefont {C.~J.}\ \bibnamefont {Palmstr\o{}m}},\ }\bibfield
  {title} {\enquote {\bibinfo {title} {{Two-dimensional epitaxial
  superconductor-semiconductor heterostructures: A platform for topological
  superconducting networks}},}\ }\href
  {https://link.aps.org/doi/10.1103/PhysRevB.93.155402} {\bibfield  {journal}
  {\bibinfo  {journal} {Phys. Rev. B}\ }\textbf {\bibinfo {volume} {93}},\
  \bibinfo {pages} {155402} (\bibinfo {year} {2016})}\BibitemShut {NoStop}%
\bibitem [{\citenamefont {Karzig}\ \emph {et~al.}(2017)\citenamefont {Karzig},
  \citenamefont {Knapp}, \citenamefont {Lutchyn}, \citenamefont {Bonderson},
  \citenamefont {Hastings}, \citenamefont {Nayak}, \citenamefont {Alicea},
  \citenamefont {Flensberg}, \citenamefont {Plugge}, \citenamefont {Oreg},
  \citenamefont {Marcus},\ and\ \citenamefont {Freedman}}]{Karzig_PRB2017}%
  \BibitemOpen
  \bibfield  {author} {\bibinfo {author} {\bibfnamefont {T.}~\bibnamefont
  {Karzig}}, \bibinfo {author} {\bibfnamefont {C.}~\bibnamefont {Knapp}},
  \bibinfo {author} {\bibfnamefont {R.~M.}\ \bibnamefont {Lutchyn}}, \bibinfo
  {author} {\bibfnamefont {P.}~\bibnamefont {Bonderson}}, \bibinfo {author}
  {\bibfnamefont {M.~B.}\ \bibnamefont {Hastings}}, \bibinfo {author}
  {\bibfnamefont {C.}~\bibnamefont {Nayak}}, \bibinfo {author} {\bibfnamefont
  {J.}~\bibnamefont {Alicea}}, \bibinfo {author} {\bibfnamefont
  {K.}~\bibnamefont {Flensberg}}, \bibinfo {author} {\bibfnamefont
  {S.}~\bibnamefont {Plugge}}, \bibinfo {author} {\bibfnamefont
  {Y.}~\bibnamefont {Oreg}}, \bibinfo {author} {\bibfnamefont {C.~M.}\
  \bibnamefont {Marcus}}, \ and\ \bibinfo {author} {\bibfnamefont {M.~H.}\
  \bibnamefont {Freedman}},\ }\bibfield  {title} {\enquote {\bibinfo {title}
  {{Scalable designs for quasiparticle-poisoning-protected topological quantum
  computation with Majorana zero modes}},}\ }\href
  {https://link.aps.org/doi/10.1103/PhysRevB.95.235305} {\bibfield  {journal}
  {\bibinfo  {journal} {Phys. Rev. B}\ }\textbf {\bibinfo {volume} {95}},\
  \bibinfo {pages} {235305} (\bibinfo {year} {2017})}\BibitemShut {NoStop}%
\bibitem [{\citenamefont {Pientka}\ \emph {et~al.}(2017)\citenamefont
  {Pientka}, \citenamefont {Keselman}, \citenamefont {Berg}, \citenamefont
  {Yacoby}, \citenamefont {Stern},\ and\ \citenamefont
  {Halperin}}]{Pientka_PRX2017}%
  \BibitemOpen
  \bibfield  {author} {\bibinfo {author} {\bibfnamefont {F.}~\bibnamefont
  {Pientka}}, \bibinfo {author} {\bibfnamefont {A.}~\bibnamefont {Keselman}},
  \bibinfo {author} {\bibfnamefont {E.}~\bibnamefont {Berg}}, \bibinfo {author}
  {\bibfnamefont {A.}~\bibnamefont {Yacoby}}, \bibinfo {author} {\bibfnamefont
  {A.}~\bibnamefont {Stern}}, \ and\ \bibinfo {author} {\bibfnamefont {B.~I.}\
  \bibnamefont {Halperin}},\ }\bibfield  {title} {\enquote {\bibinfo {title}
  {{Topological Superconductivity in a Planar Josephson Junction}},}\ }\href
  {https://link.aps.org/doi/10.1103/PhysRevX.7.021032} {\bibfield  {journal}
  {\bibinfo  {journal} {Phys. Rev. X}\ }\textbf {\bibinfo {volume} {7}},\
  \bibinfo {pages} {021032} (\bibinfo {year} {2017})}\BibitemShut {NoStop}%
\bibitem [{\citenamefont {Ren}\ \emph {et~al.}(2019)\citenamefont {Ren},
  \citenamefont {Pientka}, \citenamefont {Hart}, \citenamefont {Pierce},
  \citenamefont {Kosowsky}, \citenamefont {Lunczer}, \citenamefont {Schlereth},
  \citenamefont {Scharf}, \citenamefont {Hankiewicz}, \citenamefont
  {Molenkamp}, \citenamefont {Halperin},\ and\ \citenamefont
  {Yacoby}}]{Ren_Nature2019}%
  \BibitemOpen
  \bibfield  {author} {\bibinfo {author} {\bibfnamefont {H.}~\bibnamefont
  {Ren}}, \bibinfo {author} {\bibfnamefont {F.}~\bibnamefont {Pientka}},
  \bibinfo {author} {\bibfnamefont {S.}~\bibnamefont {Hart}}, \bibinfo {author}
  {\bibfnamefont {A.~T.}\ \bibnamefont {Pierce}}, \bibinfo {author}
  {\bibfnamefont {M.}~\bibnamefont {Kosowsky}}, \bibinfo {author}
  {\bibfnamefont {L.}~\bibnamefont {Lunczer}}, \bibinfo {author} {\bibfnamefont
  {R.}~\bibnamefont {Schlereth}}, \bibinfo {author} {\bibfnamefont
  {B.}~\bibnamefont {Scharf}}, \bibinfo {author} {\bibfnamefont {E.~M.}\
  \bibnamefont {Hankiewicz}}, \bibinfo {author} {\bibfnamefont {L.~W.}\
  \bibnamefont {Molenkamp}}, \bibinfo {author} {\bibfnamefont {B.~I.}\
  \bibnamefont {Halperin}}, \ and\ \bibinfo {author} {\bibfnamefont
  {A.}~\bibnamefont {Yacoby}},\ }\bibfield  {title} {\enquote {\bibinfo {title}
  {{Topological superconductivity in a phase-controlled Josephson junction}},}\
  }\href {https://doi.org/10.1038/s41586-019-1148-9} {\bibfield  {journal}
  {\bibinfo  {journal} {Nature}\ }\textbf {\bibinfo {volume} {569}},\ \bibinfo
  {pages} {93} (\bibinfo {year} {2019})}\BibitemShut {NoStop}%
\bibitem [{\citenamefont {Fornieri}\ \emph {et~al.}(2019)\citenamefont
  {Fornieri}, \citenamefont {Whiticar}, \citenamefont {Setiawan}, \citenamefont
  {Portol{\'e}s}, \citenamefont {Drachmann}, \citenamefont {Keselman},
  \citenamefont {Gronin}, \citenamefont {Thomas}, \citenamefont {Wang},
  \citenamefont {Kallaher}, \citenamefont {Gardner}, \citenamefont {Berg},
  \citenamefont {Manfra}, \citenamefont {Stern}, \citenamefont {Marcus},\ and\
  \citenamefont {Nichele}}]{Fornieri_Nature2019}%
  \BibitemOpen
  \bibfield  {author} {\bibinfo {author} {\bibfnamefont {A.}~\bibnamefont
  {Fornieri}}, \bibinfo {author} {\bibfnamefont {A.~M.}\ \bibnamefont
  {Whiticar}}, \bibinfo {author} {\bibfnamefont {F.}~\bibnamefont {Setiawan}},
  \bibinfo {author} {\bibfnamefont {E.}~\bibnamefont {Portol{\'e}s}}, \bibinfo
  {author} {\bibfnamefont {A.~C.~C.}\ \bibnamefont {Drachmann}}, \bibinfo
  {author} {\bibfnamefont {A.}~\bibnamefont {Keselman}}, \bibinfo {author}
  {\bibfnamefont {S.}~\bibnamefont {Gronin}}, \bibinfo {author} {\bibfnamefont
  {C.}~\bibnamefont {Thomas}}, \bibinfo {author} {\bibfnamefont
  {T.}~\bibnamefont {Wang}}, \bibinfo {author} {\bibfnamefont {R.}~\bibnamefont
  {Kallaher}}, \bibinfo {author} {\bibfnamefont {G.~C.}\ \bibnamefont
  {Gardner}}, \bibinfo {author} {\bibfnamefont {E.}~\bibnamefont {Berg}},
  \bibinfo {author} {\bibfnamefont {M.~J.}\ \bibnamefont {Manfra}}, \bibinfo
  {author} {\bibfnamefont {A.}~\bibnamefont {Stern}}, \bibinfo {author}
  {\bibfnamefont {C.~M.}\ \bibnamefont {Marcus}}, \ and\ \bibinfo {author}
  {\bibfnamefont {F.}~\bibnamefont {Nichele}},\ }\bibfield  {title} {\enquote
  {\bibinfo {title} {{Evidence of topological superconductivity in planar
  Josephson junctions}},}\ }\href {https://doi.org/10.1038/s41586-019-1068-8}
  {\bibfield  {journal} {\bibinfo  {journal} {Nature}\ }\textbf {\bibinfo
  {volume} {569}},\ \bibinfo {pages} {89} (\bibinfo {year} {2019})}\BibitemShut
  {NoStop}%
\bibitem [{\citenamefont {Dartiailh}\ \emph {et~al.}(2021)\citenamefont
  {Dartiailh}, \citenamefont {Mayer}, \citenamefont {Yuan}, \citenamefont
  {Wickramasinghe}, \citenamefont {Matos-Abiague}, \citenamefont {\ifmmode
  \check{Z}\else \v{Z}\fi{}uti\ifmmode~\acute{c}\else \'{c}\fi{}},\ and\
  \citenamefont {Shabani}}]{Javad_PRL2021}%
  \BibitemOpen
  \bibfield  {author} {\bibinfo {author} {\bibfnamefont {M.~C.}\ \bibnamefont
  {Dartiailh}}, \bibinfo {author} {\bibfnamefont {W.}~\bibnamefont {Mayer}},
  \bibinfo {author} {\bibfnamefont {J.}~\bibnamefont {Yuan}}, \bibinfo {author}
  {\bibfnamefont {K.~S.}\ \bibnamefont {Wickramasinghe}}, \bibinfo {author}
  {\bibfnamefont {A.}~\bibnamefont {Matos-Abiague}}, \bibinfo {author}
  {\bibfnamefont {I.}~\bibnamefont {\ifmmode \check{Z}\else
  \v{Z}\fi{}uti\ifmmode~\acute{c}\else \'{c}\fi{}}}, \ and\ \bibinfo {author}
  {\bibfnamefont {J.}~\bibnamefont {Shabani}},\ }\bibfield  {title} {\enquote
  {\bibinfo {title} {Phase signature of topological transition in {J}osephson
  junctions},}\ }\href
  {https://link.aps.org/doi/10.1103/PhysRevLett.126.036802} {\bibfield
  {journal} {\bibinfo  {journal} {Phys. Rev. Lett.}\ }\textbf {\bibinfo
  {volume} {126}},\ \bibinfo {pages} {036802} (\bibinfo {year}
  {2021})}\BibitemShut {NoStop}%
\bibitem [{\citenamefont {Zhou}\ \emph {et~al.}(2020)\citenamefont {Zhou},
  \citenamefont {Dartiailh}, \citenamefont {Mayer}, \citenamefont {Han},
  \citenamefont {Matos-Abiague}, \citenamefont {Shabani},\ and\ \citenamefont
  {\ifmmode \check{Z}\else \v{Z}\fi{}uti\ifmmode~\acute{c}\else
  \'{c}\fi{}}}]{Tong_PRL2020}%
  \BibitemOpen
  \bibfield  {author} {\bibinfo {author} {\bibfnamefont {T.}~\bibnamefont
  {Zhou}}, \bibinfo {author} {\bibfnamefont {M.~C.}\ \bibnamefont {Dartiailh}},
  \bibinfo {author} {\bibfnamefont {W.}~\bibnamefont {Mayer}}, \bibinfo
  {author} {\bibfnamefont {J.~E.}\ \bibnamefont {Han}}, \bibinfo {author}
  {\bibfnamefont {A.}~\bibnamefont {Matos-Abiague}}, \bibinfo {author}
  {\bibfnamefont {J.}~\bibnamefont {Shabani}}, \ and\ \bibinfo {author}
  {\bibfnamefont {I.}~\bibnamefont {\ifmmode \check{Z}\else
  \v{Z}\fi{}uti\ifmmode~\acute{c}\else \'{c}\fi{}}},\ }\bibfield  {title}
  {\enquote {\bibinfo {title} {Phase control of {M}ajorana bound states in a
  topological $\mathsf{X}$ junction},}\ }\href
  {https://link.aps.org/doi/10.1103/PhysRevLett.124.137001} {\bibfield
  {journal} {\bibinfo  {journal} {Phys. Rev. Lett.}\ }\textbf {\bibinfo
  {volume} {124}},\ \bibinfo {pages} {137001} (\bibinfo {year}
  {2020})}\BibitemShut {NoStop}%
\bibitem [{\citenamefont {Alidoust}\ \emph {et~al.}(2018)\citenamefont
  {Alidoust}, \citenamefont {Willatzen},\ and\ \citenamefont
  {Jauho}}]{Alidoust_PRB2018}%
  \BibitemOpen
  \bibfield  {author} {\bibinfo {author} {\bibfnamefont {M.}~\bibnamefont
  {Alidoust}}, \bibinfo {author} {\bibfnamefont {M.}~\bibnamefont {Willatzen}},
  \ and\ \bibinfo {author} {\bibfnamefont {A.-P.}\ \bibnamefont {Jauho}},\
  }\bibfield  {title} {\enquote {\bibinfo {title} {Strain-engineered {M}ajorana
  zero energy modes and ${\ensuremath{\varphi}}_{0}$ {J}osephson state in black
  phosphorus},}\ }\href {https://link.aps.org/doi/10.1103/PhysRevB.98.085414}
  {\bibfield  {journal} {\bibinfo  {journal} {Phys. Rev. B}\ }\textbf {\bibinfo
  {volume} {98}},\ \bibinfo {pages} {085414} (\bibinfo {year}
  {2018})}\BibitemShut {NoStop}%
\bibitem [{\citenamefont {Alidoust}\ \emph {et~al.}(2021)\citenamefont
  {Alidoust}, \citenamefont {Shen},\ and\ \citenamefont {\ifmmode
  \check{Z}\else \v{Z}\fi{}uti\ifmmode~\acute{c}\else
  \'{c}\fi{}}}]{Alidoust_PRB2021}%
  \BibitemOpen
  \bibfield  {author} {\bibinfo {author} {\bibfnamefont {M.}~\bibnamefont
  {Alidoust}}, \bibinfo {author} {\bibfnamefont {C.}~\bibnamefont {Shen}}, \
  and\ \bibinfo {author} {\bibfnamefont {I.}~\bibnamefont {\ifmmode
  \check{Z}\else \v{Z}\fi{}uti\ifmmode~\acute{c}\else \'{c}\fi{}}},\ }\bibfield
   {title} {\enquote {\bibinfo {title} {Cubic spin-orbit coupling and anomalous
  {J}osephson effect in planar junctions},}\ }\href
  {https://link.aps.org/doi/10.1103/PhysRevB.103.L060503} {\bibfield  {journal}
  {\bibinfo  {journal} {Phys. Rev. B}\ }\textbf {\bibinfo {volume} {103}},\
  \bibinfo {pages} {L060503} (\bibinfo {year} {2021})}\BibitemShut {NoStop}%
\bibitem [{\citenamefont {Setiawan}\ \emph {et~al.}(2019)\citenamefont
  {Setiawan}, \citenamefont {Stern},\ and\ \citenamefont
  {Berg}}]{Setiawan_PRB2019_1}%
  \BibitemOpen
  \bibfield  {author} {\bibinfo {author} {\bibfnamefont {F.}~\bibnamefont
  {Setiawan}}, \bibinfo {author} {\bibfnamefont {A.}~\bibnamefont {Stern}}, \
  and\ \bibinfo {author} {\bibfnamefont {E.}~\bibnamefont {Berg}},\ }\bibfield
  {title} {\enquote {\bibinfo {title} {Topological superconductivity in planar
  {J}osephson junctions: {N}arrowing down to the nanowire limit},}\ }\href
  {https://link.aps.org/doi/10.1103/PhysRevB.99.220506} {\bibfield  {journal}
  {\bibinfo  {journal} {Phys. Rev. B}\ }\textbf {\bibinfo {volume} {99}},\
  \bibinfo {pages} {220506} (\bibinfo {year} {2019})}\BibitemShut {NoStop}%
\bibitem [{\citenamefont {Pakizer}\ \emph {et~al.}(2021)\citenamefont
  {Pakizer}, \citenamefont {Scharf},\ and\ \citenamefont
  {Matos-Abiague}}]{Alex_PRRes2021}%
  \BibitemOpen
  \bibfield  {author} {\bibinfo {author} {\bibfnamefont {J.~D.}\ \bibnamefont
  {Pakizer}}, \bibinfo {author} {\bibfnamefont {B.}~\bibnamefont {Scharf}}, \
  and\ \bibinfo {author} {\bibfnamefont {A.}~\bibnamefont {Matos-Abiague}},\
  }\bibfield  {title} {\enquote {\bibinfo {title} {Crystalline anisotropic
  topological superconductivity in planar {J}osephson junctions},}\ }\href
  {\doibase 10.1103/PhysRevResearch.3.013198} {\bibfield  {journal} {\bibinfo
  {journal} {Phys. Rev. Research}\ }\textbf {\bibinfo {volume} {3}},\ \bibinfo
  {pages} {013198} (\bibinfo {year} {2021})}\BibitemShut {NoStop}%
\bibitem [{\citenamefont {Fu}\ and\ \citenamefont {Kane}(2008)}]{Fu_PRL2008}%
  \BibitemOpen
  \bibfield  {author} {\bibinfo {author} {\bibfnamefont {L.}~\bibnamefont
  {Fu}}\ and\ \bibinfo {author} {\bibfnamefont {C.~L.}\ \bibnamefont {Kane}},\
  }\bibfield  {title} {\enquote {\bibinfo {title} {Superconducting proximity
  effect and {M}ajorana fermions at the surface of a topological insulator},}\
  }\href {https://link.aps.org/doi/10.1103/PhysRevLett.100.096407} {\bibfield
  {journal} {\bibinfo  {journal} {Phys. Rev. Lett.}\ }\textbf {\bibinfo
  {volume} {100}},\ \bibinfo {pages} {096407} (\bibinfo {year}
  {2008})}\BibitemShut {NoStop}%
\bibitem [{\citenamefont {Yang}\ \emph {et~al.}(2016)\citenamefont {Yang},
  \citenamefont {Stano}, \citenamefont {Klinovaja},\ and\ \citenamefont
  {Loss}}]{Loss_PRB2016}%
  \BibitemOpen
  \bibfield  {author} {\bibinfo {author} {\bibfnamefont {G.}~\bibnamefont
  {Yang}}, \bibinfo {author} {\bibfnamefont {P.}~\bibnamefont {Stano}},
  \bibinfo {author} {\bibfnamefont {J.}~\bibnamefont {Klinovaja}}, \ and\
  \bibinfo {author} {\bibfnamefont {D.}~\bibnamefont {Loss}},\ }\bibfield
  {title} {\enquote {\bibinfo {title} {Majorana bound states in magnetic
  skyrmions},}\ }\href {https://link.aps.org/doi/10.1103/PhysRevB.93.224505}
  {\bibfield  {journal} {\bibinfo  {journal} {Phys. Rev. B}\ }\textbf {\bibinfo
  {volume} {93}},\ \bibinfo {pages} {224505} (\bibinfo {year}
  {2016})}\BibitemShut {NoStop}%
\bibitem [{\citenamefont {G\"ung\"ord\"u}\ \emph {et~al.}(2018)\citenamefont
  {G\"ung\"ord\"u}, \citenamefont {Sandhoefner},\ and\ \citenamefont
  {Kovalev}}]{Kovalev_PRB2018}%
  \BibitemOpen
  \bibfield  {author} {\bibinfo {author} {\bibfnamefont {U.}~\bibnamefont
  {G\"ung\"ord\"u}}, \bibinfo {author} {\bibfnamefont {S.}~\bibnamefont
  {Sandhoefner}}, \ and\ \bibinfo {author} {\bibfnamefont {A.~A.}\ \bibnamefont
  {Kovalev}},\ }\bibfield  {title} {\enquote {\bibinfo {title} {Stabilization
  and control of {M}ajorana bound states with elongated skyrmions},}\ }\href
  {https://link.aps.org/doi/10.1103/PhysRevB.97.115136} {\bibfield  {journal}
  {\bibinfo  {journal} {Phys. Rev. B}\ }\textbf {\bibinfo {volume} {97}},\
  \bibinfo {pages} {115136} (\bibinfo {year} {2018})}\BibitemShut {NoStop}%
\bibitem [{\citenamefont {Garnier}\ \emph {et~al.}(2019)\citenamefont
  {Garnier}, \citenamefont {Mesaros},\ and\ \citenamefont
  {Simon}}]{Garnier_CommunPhys2019}%
  \BibitemOpen
  \bibfield  {author} {\bibinfo {author} {\bibfnamefont {M.}~\bibnamefont
  {Garnier}}, \bibinfo {author} {\bibfnamefont {A.}~\bibnamefont {Mesaros}}, \
  and\ \bibinfo {author} {\bibfnamefont {P.}~\bibnamefont {Simon}},\ }\bibfield
   {title} {\enquote {\bibinfo {title} {Topological superconductivity with
  deformable magnetic skyrmions},}\ }\href
  {https://doi.org/10.1038/s42005-019-0226-5} {\bibfield  {journal} {\bibinfo
  {journal} {Commun. Phys.}\ }\textbf {\bibinfo {volume} {2}},\ \bibinfo
  {pages} {126} (\bibinfo {year} {2019})}\BibitemShut {NoStop}%
\bibitem [{\citenamefont {Mohanta}\ \emph
  {et~al.}(2020{\natexlab{a}})\citenamefont {Mohanta}, \citenamefont
  {Taraphder}, \citenamefont {Dagotto},\ and\ \citenamefont
  {Okamoto}}]{Mohanta_PRB2020}%
  \BibitemOpen
  \bibfield  {author} {\bibinfo {author} {\bibfnamefont {N.}~\bibnamefont
  {Mohanta}}, \bibinfo {author} {\bibfnamefont {A.}~\bibnamefont {Taraphder}},
  \bibinfo {author} {\bibfnamefont {E.}~\bibnamefont {Dagotto}}, \ and\
  \bibinfo {author} {\bibfnamefont {S.}~\bibnamefont {Okamoto}},\ }\bibfield
  {title} {\enquote {\bibinfo {title} {Magnetic switching in {W}eyl
  semimetal-superconductor heterostructures},}\ }\href
  {https://link.aps.org/doi/10.1103/PhysRevB.102.064506} {\bibfield  {journal}
  {\bibinfo  {journal} {Phys. Rev. B}\ }\textbf {\bibinfo {volume} {102}},\
  \bibinfo {pages} {064506} (\bibinfo {year} {2020}{\natexlab{a}})}\BibitemShut
  {NoStop}%
\bibitem [{\citenamefont {Mascot}\ \emph {et~al.}(2021)\citenamefont {Mascot},
  \citenamefont {Bedow}, \citenamefont {Graham}, \citenamefont {Rachel},\ and\
  \citenamefont {Morr}}]{Mascot_npjQM2021}%
  \BibitemOpen
  \bibfield  {author} {\bibinfo {author} {\bibfnamefont {E.}~\bibnamefont
  {Mascot}}, \bibinfo {author} {\bibfnamefont {J.}~\bibnamefont {Bedow}},
  \bibinfo {author} {\bibfnamefont {M.}~\bibnamefont {Graham}}, \bibinfo
  {author} {\bibfnamefont {S.}~\bibnamefont {Rachel}}, \ and\ \bibinfo {author}
  {\bibfnamefont {D.~K.}\ \bibnamefont {Morr}},\ }\bibfield  {title} {\enquote
  {\bibinfo {title} {Topological superconductivity in skyrmion lattices},}\
  }\href {https://doi.org/10.1038/s41535-020-00299-x} {\bibfield  {journal}
  {\bibinfo  {journal} {npj Quantum Mater.}\ }\textbf {\bibinfo {volume} {6}},\
  \bibinfo {pages} {6} (\bibinfo {year} {2021})}\BibitemShut {NoStop}%
\bibitem [{\citenamefont {Bedow}\ \emph {et~al.}(2020)\citenamefont {Bedow},
  \citenamefont {Mascot}, \citenamefont {Posske}, \citenamefont {Uhrig},
  \citenamefont {Wiesendanger}, \citenamefont {Rachel},\ and\ \citenamefont
  {Morr}}]{Bedow_PRB2020}%
  \BibitemOpen
  \bibfield  {author} {\bibinfo {author} {\bibfnamefont {J.}~\bibnamefont
  {Bedow}}, \bibinfo {author} {\bibfnamefont {E.}~\bibnamefont {Mascot}},
  \bibinfo {author} {\bibfnamefont {T.}~\bibnamefont {Posske}}, \bibinfo
  {author} {\bibfnamefont {G.~S.}\ \bibnamefont {Uhrig}}, \bibinfo {author}
  {\bibfnamefont {R.}~\bibnamefont {Wiesendanger}}, \bibinfo {author}
  {\bibfnamefont {S.}~\bibnamefont {Rachel}}, \ and\ \bibinfo {author}
  {\bibfnamefont {D.~K.}\ \bibnamefont {Morr}},\ }\bibfield  {title} {\enquote
  {\bibinfo {title} {Topological superconductivity induced by a triple-q
  magnetic structure},}\ }\href
  {https://link.aps.org/doi/10.1103/PhysRevB.102.180504} {\bibfield  {journal}
  {\bibinfo  {journal} {Phys. Rev. B}\ }\textbf {\bibinfo {volume} {102}},\
  \bibinfo {pages} {180504} (\bibinfo {year} {2020})}\BibitemShut {NoStop}%
\bibitem [{\citenamefont {Rex}\ \emph {et~al.}(2019)\citenamefont {Rex},
  \citenamefont {Gornyi},\ and\ \citenamefont {Mirlin}}]{Rex_PRB2019}%
  \BibitemOpen
  \bibfield  {author} {\bibinfo {author} {\bibfnamefont {S.}~\bibnamefont
  {Rex}}, \bibinfo {author} {\bibfnamefont {I.~V.}\ \bibnamefont {Gornyi}}, \
  and\ \bibinfo {author} {\bibfnamefont {A.~D.}\ \bibnamefont {Mirlin}},\
  }\bibfield  {title} {\enquote {\bibinfo {title} {Majorana bound states in
  magnetic skyrmions imposed onto a superconductor},}\ }\href
  {https://link.aps.org/doi/10.1103/PhysRevB.100.064504} {\bibfield  {journal}
  {\bibinfo  {journal} {Phys. Rev. B}\ }\textbf {\bibinfo {volume} {100}},\
  \bibinfo {pages} {064504} (\bibinfo {year} {2019})}\BibitemShut {NoStop}%
\bibitem [{\citenamefont {Rex}\ \emph {et~al.}(2020)\citenamefont {Rex},
  \citenamefont {Gornyi},\ and\ \citenamefont {Mirlin}}]{Rex_PRB2020}%
  \BibitemOpen
  \bibfield  {author} {\bibinfo {author} {\bibfnamefont {S.}~\bibnamefont
  {Rex}}, \bibinfo {author} {\bibfnamefont {I.~V.}\ \bibnamefont {Gornyi}}, \
  and\ \bibinfo {author} {\bibfnamefont {A.~D.}\ \bibnamefont {Mirlin}},\
  }\bibfield  {title} {\enquote {\bibinfo {title} {Majorana modes in
  emergent-wire phases of helical and cycloidal magnet-superconductor
  hybrids},}\ }\href {https://link.aps.org/doi/10.1103/PhysRevB.102.224501}
  {\bibfield  {journal} {\bibinfo  {journal} {Phys. Rev. B}\ }\textbf {\bibinfo
  {volume} {102}},\ \bibinfo {pages} {224501} (\bibinfo {year}
  {2020})}\BibitemShut {NoStop}%
\bibitem [{\citenamefont {Mohanta}\ \emph
  {et~al.}(2019{\natexlab{b}})\citenamefont {Mohanta}, \citenamefont
  {Dagotto},\ and\ \citenamefont {Okamoto}}]{NM_PRB2019}%
  \BibitemOpen
  \bibfield  {author} {\bibinfo {author} {\bibfnamefont {N.}~\bibnamefont
  {Mohanta}}, \bibinfo {author} {\bibfnamefont {E.}~\bibnamefont {Dagotto}}, \
  and\ \bibinfo {author} {\bibfnamefont {S.}~\bibnamefont {Okamoto}},\
  }\bibfield  {title} {\enquote {\bibinfo {title} {Topological {H}all effect
  and emergent skyrmion crystal at manganite-iridate oxide interfaces},}\
  }\href {https://link.aps.org/doi/10.1103/PhysRevB.100.064429} {\bibfield
  {journal} {\bibinfo  {journal} {Phys. Rev. B}\ }\textbf {\bibinfo {volume}
  {100}},\ \bibinfo {pages} {064429} (\bibinfo {year}
  {2019}{\natexlab{b}})}\BibitemShut {NoStop}%
\bibitem [{\citenamefont {Mohanta}\ \emph
  {et~al.}(2020{\natexlab{b}})\citenamefont {Mohanta}, \citenamefont
  {Okamoto},\ and\ \citenamefont {Dagotto}}]{NM_PRB2020}%
  \BibitemOpen
  \bibfield  {author} {\bibinfo {author} {\bibfnamefont {N.}~\bibnamefont
  {Mohanta}}, \bibinfo {author} {\bibfnamefont {S.}~\bibnamefont {Okamoto}}, \
  and\ \bibinfo {author} {\bibfnamefont {E.}~\bibnamefont {Dagotto}},\
  }\bibfield  {title} {\enquote {\bibinfo {title} {{Planar topological Hall
  effect from conical spin spirals}},}\ }\href
  {https://link.aps.org/doi/10.1103/PhysRevB.102.064430} {\bibfield  {journal}
  {\bibinfo  {journal} {Phys. Rev. B}\ }\textbf {\bibinfo {volume} {102}},\
  \bibinfo {pages} {064430} (\bibinfo {year} {2020}{\natexlab{b}})}\BibitemShut
  {NoStop}%
\bibitem [{\citenamefont {Mohanta}\ \emph
  {et~al.}(2020{\natexlab{c}})\citenamefont {Mohanta}, \citenamefont
  {Christianson}, \citenamefont {Okamoto},\ and\ \citenamefont
  {Dagotto}}]{Mohanta_CommunPhys2020}%
  \BibitemOpen
  \bibfield  {author} {\bibinfo {author} {\bibfnamefont {N.}~\bibnamefont
  {Mohanta}}, \bibinfo {author} {\bibfnamefont {A.~D.}\ \bibnamefont
  {Christianson}}, \bibinfo {author} {\bibfnamefont {S.}~\bibnamefont
  {Okamoto}}, \ and\ \bibinfo {author} {\bibfnamefont {E.}~\bibnamefont
  {Dagotto}},\ }\bibfield  {title} {\enquote {\bibinfo {title} {Signatures of a
  liquid-crystal transition in spin-wave excitations of skyrmions},}\ }\href
  {https://doi.org/10.1038/s42005-020-00489-w} {\bibfield  {journal} {\bibinfo
  {journal} {Commun. Phys.}\ }\textbf {\bibinfo {volume} {3}},\ \bibinfo
  {pages} {229} (\bibinfo {year} {2020}{\natexlab{c}})}\BibitemShut {NoStop}%
\bibitem [{\citenamefont {R{\"o}{\ss}ler}\ \emph {et~al.}(2006)\citenamefont
  {R{\"o}{\ss}ler}, \citenamefont {Bogdanov},\ and\ \citenamefont
  {Pfleiderer}}]{Rossler_Nature2006}%
  \BibitemOpen
  \bibfield  {author} {\bibinfo {author} {\bibfnamefont {U.~K.}\ \bibnamefont
  {R{\"o}{\ss}ler}}, \bibinfo {author} {\bibfnamefont {A.~N.}\ \bibnamefont
  {Bogdanov}}, \ and\ \bibinfo {author} {\bibfnamefont {C.}~\bibnamefont
  {Pfleiderer}},\ }\bibfield  {title} {\enquote {\bibinfo {title} {Spontaneous
  skyrmion ground states in magnetic metals},}\ }\href
  {https://doi.org/10.1038/nature05056} {\bibfield  {journal} {\bibinfo
  {journal} {Nature}\ }\textbf {\bibinfo {volume} {442}},\ \bibinfo {pages}
  {797} (\bibinfo {year} {2006})}\BibitemShut {NoStop}%
\bibitem [{\citenamefont {M{\"u}hlbauer}\ \emph {et~al.}(2009)\citenamefont
  {M{\"u}hlbauer}, \citenamefont {Binz}, \citenamefont {Jonietz}, \citenamefont
  {Pfleiderer}, \citenamefont {Rosch}, \citenamefont {Neubauer}, \citenamefont
  {Georgii},\ and\ \citenamefont {B{\"o}ni}}]{Muhlbauer_Science2009}%
  \BibitemOpen
  \bibfield  {author} {\bibinfo {author} {\bibfnamefont {S.}~\bibnamefont
  {M{\"u}hlbauer}}, \bibinfo {author} {\bibfnamefont {B.}~\bibnamefont {Binz}},
  \bibinfo {author} {\bibfnamefont {F.}~\bibnamefont {Jonietz}}, \bibinfo
  {author} {\bibfnamefont {C.}~\bibnamefont {Pfleiderer}}, \bibinfo {author}
  {\bibfnamefont {A.}~\bibnamefont {Rosch}}, \bibinfo {author} {\bibfnamefont
  {A.}~\bibnamefont {Neubauer}}, \bibinfo {author} {\bibfnamefont
  {R.}~\bibnamefont {Georgii}}, \ and\ \bibinfo {author} {\bibfnamefont
  {P.}~\bibnamefont {B{\"o}ni}},\ }\bibfield  {title} {\enquote {\bibinfo
  {title} {Skyrmion lattice in a chiral magnet},}\ }\href
  {https://science.sciencemag.org/content/323/5916/915} {\bibfield  {journal}
  {\bibinfo  {journal} {Science}\ }\textbf {\bibinfo {volume} {323}},\ \bibinfo
  {pages} {915} (\bibinfo {year} {2009})}\BibitemShut {NoStop}%
\bibitem [{\citenamefont {Yu}\ \emph {et~al.}(2010)\citenamefont {Yu},
  \citenamefont {Onose}, \citenamefont {Kanazawa}, \citenamefont {Park},
  \citenamefont {Han}, \citenamefont {Matsui}, \citenamefont {Nagaosa},\ and\
  \citenamefont {Tokura}}]{Yu_Nature2010}%
  \BibitemOpen
  \bibfield  {author} {\bibinfo {author} {\bibfnamefont {X.~Z.}\ \bibnamefont
  {Yu}}, \bibinfo {author} {\bibfnamefont {Y.}~\bibnamefont {Onose}}, \bibinfo
  {author} {\bibfnamefont {N.}~\bibnamefont {Kanazawa}}, \bibinfo {author}
  {\bibfnamefont {J.~H.}\ \bibnamefont {Park}}, \bibinfo {author}
  {\bibfnamefont {J.~H.}\ \bibnamefont {Han}}, \bibinfo {author} {\bibfnamefont
  {Y.}~\bibnamefont {Matsui}}, \bibinfo {author} {\bibfnamefont
  {N.}~\bibnamefont {Nagaosa}}, \ and\ \bibinfo {author} {\bibfnamefont
  {Y.}~\bibnamefont {Tokura}},\ }\bibfield  {title} {\enquote {\bibinfo {title}
  {Real-space observation of a two-dimensional skyrmion crystal},}\ }\href
  {https://doi.org/10.1038/nature09124} {\bibfield  {journal} {\bibinfo
  {journal} {Nature}\ }\textbf {\bibinfo {volume} {465}},\ \bibinfo {pages}
  {901} (\bibinfo {year} {2010})}\BibitemShut {NoStop}%
\bibitem [{\citenamefont {Mohanta}\ \emph {et~al.}(2017)\citenamefont
  {Mohanta}, \citenamefont {Kampf},\ and\ \citenamefont
  {Kopp}}]{Mohanta_SciRep2017}%
  \BibitemOpen
  \bibfield  {author} {\bibinfo {author} {\bibfnamefont {N.}~\bibnamefont
  {Mohanta}}, \bibinfo {author} {\bibfnamefont {A.~P.}\ \bibnamefont {Kampf}},
  \ and\ \bibinfo {author} {\bibfnamefont {T.}~\bibnamefont {Kopp}},\
  }\bibfield  {title} {\enquote {\bibinfo {title} {Emergent momentum-space
  {S}kyrmion texture on the surface of topological insulators},}\ }\href
  {https://doi.org/10.1038/srep45664} {\bibfield  {journal} {\bibinfo
  {journal} {Sci. Rep.}\ }\textbf {\bibinfo {volume} {7}},\ \bibinfo {pages}
  {45664} (\bibinfo {year} {2017})}\BibitemShut {NoStop}%
\bibitem [{\citenamefont {Sun}\ \emph {et~al.}(2013)\citenamefont {Sun},
  \citenamefont {Cao}, \citenamefont {Miao}, \citenamefont {Feng},
  \citenamefont {You}, \citenamefont {Wu}, \citenamefont {Zhang}, \citenamefont
  {Hu},\ and\ \citenamefont {Ding}}]{Sun_PRL2013}%
  \BibitemOpen
  \bibfield  {author} {\bibinfo {author} {\bibfnamefont {L.}~\bibnamefont
  {Sun}}, \bibinfo {author} {\bibfnamefont {R.~X.}\ \bibnamefont {Cao}},
  \bibinfo {author} {\bibfnamefont {B.~F.}\ \bibnamefont {Miao}}, \bibinfo
  {author} {\bibfnamefont {Z.}~\bibnamefont {Feng}}, \bibinfo {author}
  {\bibfnamefont {B.}~\bibnamefont {You}}, \bibinfo {author} {\bibfnamefont
  {D.}~\bibnamefont {Wu}}, \bibinfo {author} {\bibfnamefont {W.}~\bibnamefont
  {Zhang}}, \bibinfo {author} {\bibfnamefont {An}~\bibnamefont {Hu}}, \ and\
  \bibinfo {author} {\bibfnamefont {H.~F.}\ \bibnamefont {Ding}},\ }\bibfield
  {title} {\enquote {\bibinfo {title} {Creating an artificial two-dimensional
  {S}kyrmion crystal by nanopatterning},}\ }\href
  {https://link.aps.org/doi/10.1103/PhysRevLett.110.167201} {\bibfield
  {journal} {\bibinfo  {journal} {Phys. Rev. Lett.}\ }\textbf {\bibinfo
  {volume} {110}},\ \bibinfo {pages} {167201} (\bibinfo {year}
  {2013})}\BibitemShut {NoStop}%
\bibitem [{\citenamefont {G{\"u}l}\ \emph {et~al.}(2017)\citenamefont
  {G{\"u}l}, \citenamefont {Zhang}, \citenamefont {de~Vries}, \citenamefont
  {van Veen}, \citenamefont {Zuo}, \citenamefont {Mourik}, \citenamefont
  {Conesa-Boj}, \citenamefont {Nowak}, \citenamefont {van Woerkom},
  \citenamefont {Quintero-P{\'e}rez}, \citenamefont {Cassidy}, \citenamefont
  {Geresdi}, \citenamefont {Koelling}, \citenamefont {Car}, \citenamefont
  {Plissard}, \citenamefont {Bakkers},\ and\ \citenamefont
  {Kouwenhoven}}]{Gul_NanoLett2017}%
  \BibitemOpen
  \bibfield  {author} {\bibinfo {author} {\bibfnamefont {{\"O}.}~\bibnamefont
  {G{\"u}l}}, \bibinfo {author} {\bibfnamefont {H.}~\bibnamefont {Zhang}},
  \bibinfo {author} {\bibfnamefont {F.~K.}\ \bibnamefont {de~Vries}}, \bibinfo
  {author} {\bibfnamefont {J.}~\bibnamefont {van Veen}}, \bibinfo {author}
  {\bibfnamefont {K.}~\bibnamefont {Zuo}}, \bibinfo {author} {\bibfnamefont
  {V.}~\bibnamefont {Mourik}}, \bibinfo {author} {\bibfnamefont
  {S.}~\bibnamefont {Conesa-Boj}}, \bibinfo {author} {\bibfnamefont {M.~P.}\
  \bibnamefont {Nowak}}, \bibinfo {author} {\bibfnamefont {D.~J.}\ \bibnamefont
  {van Woerkom}}, \bibinfo {author} {\bibfnamefont {M.}~\bibnamefont
  {Quintero-P{\'e}rez}}, \bibinfo {author} {\bibfnamefont {M.~C.}\ \bibnamefont
  {Cassidy}}, \bibinfo {author} {\bibfnamefont {A.}~\bibnamefont {Geresdi}},
  \bibinfo {author} {\bibfnamefont {S.}~\bibnamefont {Koelling}}, \bibinfo
  {author} {\bibfnamefont {D.}~\bibnamefont {Car}}, \bibinfo {author}
  {\bibfnamefont {S.~R.}\ \bibnamefont {Plissard}}, \bibinfo {author}
  {\bibfnamefont {E.~P. A.~M.}\ \bibnamefont {Bakkers}}, \ and\ \bibinfo
  {author} {\bibfnamefont {L.~P.}\ \bibnamefont {Kouwenhoven}},\ }\bibfield
  {title} {\enquote {\bibinfo {title} {{Hard Superconducting Gap in {InSb}
  Nanowires}},}\ }\href {https://doi.org/10.1021/acs.nanolett.7b00540}
  {\bibfield  {journal} {\bibinfo  {journal} {Nano Lett.}\ }\textbf {\bibinfo
  {volume} {17}},\ \bibinfo {pages} {2690} (\bibinfo {year}
  {2017})}\BibitemShut {NoStop}%
\bibitem [{\citenamefont {Deng}\ \emph {et~al.}(2016)\citenamefont {Deng},
  \citenamefont {Vaitiekenas}, \citenamefont {Hansen}, \citenamefont {Danon},
  \citenamefont {Leijnse}, \citenamefont {Flensberg}, \citenamefont {Nyg{\r
  a}rd}, \citenamefont {Krogstrup},\ and\ \citenamefont
  {Marcus}}]{Deng_Science2016}%
  \BibitemOpen
  \bibfield  {author} {\bibinfo {author} {\bibfnamefont {M.~T.}\ \bibnamefont
  {Deng}}, \bibinfo {author} {\bibfnamefont {S.}~\bibnamefont {Vaitiekenas}},
  \bibinfo {author} {\bibfnamefont {E.~B.}\ \bibnamefont {Hansen}}, \bibinfo
  {author} {\bibfnamefont {J.}~\bibnamefont {Danon}}, \bibinfo {author}
  {\bibfnamefont {M.}~\bibnamefont {Leijnse}}, \bibinfo {author} {\bibfnamefont
  {K.}~\bibnamefont {Flensberg}}, \bibinfo {author} {\bibfnamefont
  {J.}~\bibnamefont {Nyg{\r a}rd}}, \bibinfo {author} {\bibfnamefont
  {P.}~\bibnamefont {Krogstrup}}, \ and\ \bibinfo {author} {\bibfnamefont
  {C.~M.}\ \bibnamefont {Marcus}},\ }\bibfield  {title} {\enquote {\bibinfo
  {title} {{Majorana bound state in a coupled quantum-dot hybrid-nanowire
  system}},}\ }\href {https://science.sciencemag.org/content/354/6319/1557}
  {\bibfield  {journal} {\bibinfo  {journal} {Science}\ }\textbf {\bibinfo
  {volume} {354}},\ \bibinfo {pages} {1557} (\bibinfo {year}
  {2016})}\BibitemShut {NoStop}%
\bibitem [{\citenamefont {Nedniyom}\ \emph {et~al.}(2009)\citenamefont
  {Nedniyom}, \citenamefont {Nicholas}, \citenamefont {Emeny}, \citenamefont
  {Buckle}, \citenamefont {Gilbertson}, \citenamefont {Buckle},\ and\
  \citenamefont {Ashley}}]{Nedniyom_PRB2009}%
  \BibitemOpen
  \bibfield  {author} {\bibinfo {author} {\bibfnamefont {B.}~\bibnamefont
  {Nedniyom}}, \bibinfo {author} {\bibfnamefont {R.~J.}\ \bibnamefont
  {Nicholas}}, \bibinfo {author} {\bibfnamefont {M.~T.}\ \bibnamefont {Emeny}},
  \bibinfo {author} {\bibfnamefont {L.}~\bibnamefont {Buckle}}, \bibinfo
  {author} {\bibfnamefont {A.~M.}\ \bibnamefont {Gilbertson}}, \bibinfo
  {author} {\bibfnamefont {P.~D.}\ \bibnamefont {Buckle}}, \ and\ \bibinfo
  {author} {\bibfnamefont {T.}~\bibnamefont {Ashley}},\ }\bibfield  {title}
  {\enquote {\bibinfo {title} {{Giant enhanced g-factors in an InSb
  two-dimensional gas}},}\ }\href
  {https://link.aps.org/doi/10.1103/PhysRevB.80.125328} {\bibfield  {journal}
  {\bibinfo  {journal} {Phys. Rev. B}\ }\textbf {\bibinfo {volume} {80}},\
  \bibinfo {pages} {125328} (\bibinfo {year} {2009})}\BibitemShut {NoStop}%
\bibitem [{\citenamefont {Qu}\ \emph {et~al.}(2016)\citenamefont {Qu},
  \citenamefont {van Veen}, \citenamefont {de~Vries}, \citenamefont {Beukman},
  \citenamefont {Wimmer}, \citenamefont {Yi}, \citenamefont {Kiselev},
  \citenamefont {Nguyen}, \citenamefont {Sokolich}, \citenamefont {Manfra},
  \citenamefont {Nichele}, \citenamefont {Marcus},\ and\ \citenamefont
  {Kouwenhoven}}]{Qu_NanoLett2016}%
  \BibitemOpen
  \bibfield  {author} {\bibinfo {author} {\bibfnamefont {F.}~\bibnamefont
  {Qu}}, \bibinfo {author} {\bibfnamefont {J.}~\bibnamefont {van Veen}},
  \bibinfo {author} {\bibfnamefont {F.~K.}\ \bibnamefont {de~Vries}}, \bibinfo
  {author} {\bibfnamefont {A.~J.~A.}\ \bibnamefont {Beukman}}, \bibinfo
  {author} {\bibfnamefont {M.}~\bibnamefont {Wimmer}}, \bibinfo {author}
  {\bibfnamefont {W.}~\bibnamefont {Yi}}, \bibinfo {author} {\bibfnamefont
  {A.~A.}\ \bibnamefont {Kiselev}}, \bibinfo {author} {\bibfnamefont {B.-M.}\
  \bibnamefont {Nguyen}}, \bibinfo {author} {\bibfnamefont {M.}~\bibnamefont
  {Sokolich}}, \bibinfo {author} {\bibfnamefont {M.~J.}\ \bibnamefont
  {Manfra}}, \bibinfo {author} {\bibfnamefont {F.}~\bibnamefont {Nichele}},
  \bibinfo {author} {\bibfnamefont {C.~M.}\ \bibnamefont {Marcus}}, \ and\
  \bibinfo {author} {\bibfnamefont {L.~P.}\ \bibnamefont {Kouwenhoven}},\
  }\bibfield  {title} {\enquote {\bibinfo {title} {{Quantized conductance and
  large g-factor anisotropy in InSb quantum point contacts}},}\ }\href
  {https://doi.org/10.1021/acs.nanolett.6b03297} {\bibfield  {journal}
  {\bibinfo  {journal} {Nano Lett.}\ }\textbf {\bibinfo {volume} {16}},\
  \bibinfo {pages} {7509} (\bibinfo {year} {2016})}\BibitemShut {NoStop}%
\bibitem [{gri()}]{gridspacing}%
  \BibitemOpen
  \href@noop {} {}\bibinfo {note} {$a$ is the lattice grid spacing used to
  discretize the kinetic energy term $\frac{p^2}{2m}$ within finite-difference
  approximation. It is immaterial as long as the lattice lengths remain
  fixed.}\BibitemShut {Stop}%
\bibitem [{\citenamefont {Zeng}\ \emph {et~al.}(2019)\citenamefont {Zeng},
  \citenamefont {Chen},\ and\ \citenamefont {Lim}}]{Zeng_APL2019}%
  \BibitemOpen
  \bibfield  {author} {\bibinfo {author} {\bibfnamefont {M.}~\bibnamefont
  {Zeng}}, \bibinfo {author} {\bibfnamefont {B.}~\bibnamefont {Chen}}, \ and\
  \bibinfo {author} {\bibfnamefont {S.~T.}\ \bibnamefont {Lim}},\ }\bibfield
  {title} {\enquote {\bibinfo {title} {Interfacial electric field and
  spin-orbitronic properties of heavy-metal/{CoFe} bilayers},}\ }\href
  {https://doi.org/10.1063/1.5043444} {\bibfield  {journal} {\bibinfo
  {journal} {Appl. Phys. Lett.}\ }\textbf {\bibinfo {volume} {114}},\ \bibinfo
  {pages} {012401} (\bibinfo {year} {2019})}\BibitemShut {NoStop}%
\bibitem [{\citenamefont {Sticlet}\ \emph {et~al.}(2012)\citenamefont
  {Sticlet}, \citenamefont {Bena},\ and\ \citenamefont
  {Simon}}]{Sticlet_PRL2012}%
  \BibitemOpen
  \bibfield  {author} {\bibinfo {author} {\bibfnamefont {D.}~\bibnamefont
  {Sticlet}}, \bibinfo {author} {\bibfnamefont {C.}~\bibnamefont {Bena}}, \
  and\ \bibinfo {author} {\bibfnamefont {P.}~\bibnamefont {Simon}},\ }\bibfield
   {title} {\enquote {\bibinfo {title} {Spin and {M}ajorana polarization in
  topological superconducting wires},}\ }\href
  {https://link.aps.org/doi/10.1103/PhysRevLett.108.096802} {\bibfield
  {journal} {\bibinfo  {journal} {Phys. Rev. Lett.}\ }\textbf {\bibinfo
  {volume} {108}},\ \bibinfo {pages} {096802} (\bibinfo {year}
  {2012})}\BibitemShut {NoStop}%
\bibitem [{\citenamefont {Sedlmayr}\ and\ \citenamefont
  {Bena}(2015)}]{Sedlmayr_PRB2015}%
  \BibitemOpen
  \bibfield  {author} {\bibinfo {author} {\bibfnamefont {N.}~\bibnamefont
  {Sedlmayr}}\ and\ \bibinfo {author} {\bibfnamefont {C.}~\bibnamefont
  {Bena}},\ }\bibfield  {title} {\enquote {\bibinfo {title} {Visualizing
  {M}ajorana bound states in one and two dimensions using the generalized
  {M}ajorana polarization},}\ }\href
  {https://link.aps.org/doi/10.1103/PhysRevB.92.115115} {\bibfield  {journal}
  {\bibinfo  {journal} {Phys. Rev. B}\ }\textbf {\bibinfo {volume} {92}},\
  \bibinfo {pages} {115115} (\bibinfo {year} {2015})}\BibitemShut {NoStop}%
\bibitem [{\citenamefont {G\l{}odzik}\ \emph {et~al.}(2020)\citenamefont
  {G\l{}odzik}, \citenamefont {Sedlmayr},\ and\ \citenamefont
  {Doma\ifmmode~\acute{n}\else \'{n}\fi{}ski}}]{Godzik_PRB2020}%
  \BibitemOpen
  \bibfield  {author} {\bibinfo {author} {\bibfnamefont {S.}~\bibnamefont
  {G\l{}odzik}}, \bibinfo {author} {\bibfnamefont {N.}~\bibnamefont
  {Sedlmayr}}, \ and\ \bibinfo {author} {\bibfnamefont {T.}~\bibnamefont
  {Doma\ifmmode~\acute{n}\else \'{n}\fi{}ski}},\ }\bibfield  {title} {\enquote
  {\bibinfo {title} {How to measure the {M}ajorana polarization of a
  topological planar josephson junction},}\ }\href
  {https://link.aps.org/doi/10.1103/PhysRevB.102.085411} {\bibfield  {journal}
  {\bibinfo  {journal} {Phys. Rev. B}\ }\textbf {\bibinfo {volume} {102}},\
  \bibinfo {pages} {085411} (\bibinfo {year} {2020})}\BibitemShut {NoStop}%
\bibitem [{\citenamefont {Scharf}\ \emph {et~al.}(2019)\citenamefont {Scharf},
  \citenamefont {Pientka}, \citenamefont {Ren}, \citenamefont {Yacoby},\ and\
  \citenamefont {Hankiewicz}}]{Scharf_PRB2019}%
  \BibitemOpen
  \bibfield  {author} {\bibinfo {author} {\bibfnamefont {B.}~\bibnamefont
  {Scharf}}, \bibinfo {author} {\bibfnamefont {F.}~\bibnamefont {Pientka}},
  \bibinfo {author} {\bibfnamefont {H.}~\bibnamefont {Ren}}, \bibinfo {author}
  {\bibfnamefont {A.}~\bibnamefont {Yacoby}}, \ and\ \bibinfo {author}
  {\bibfnamefont {E.~M.}\ \bibnamefont {Hankiewicz}},\ }\bibfield  {title}
  {\enquote {\bibinfo {title} {Tuning topological superconductivity in
  phase-controlled {J}osephson junctions with {R}ashba and {D}resselhaus
  spin-orbit coupling},}\ }\href
  {https://link.aps.org/doi/10.1103/PhysRevB.99.214503} {\bibfield  {journal}
  {\bibinfo  {journal} {Phys. Rev. B}\ }\textbf {\bibinfo {volume} {99}},\
  \bibinfo {pages} {214503} (\bibinfo {year} {2019})}\BibitemShut {NoStop}%
\bibitem [{\citenamefont {Sengupta}\ \emph {et~al.}(2001)\citenamefont
  {Sengupta}, \citenamefont {\ifmmode \check{Z}\else
  \v{Z}\fi{}uti\ifmmode~\acute{c}\else \'{c}\fi{}}, \citenamefont {Kwon},
  \citenamefont {Yakovenko},\ and\ \citenamefont
  {Das~Sarma}}]{Sengupta_PRB2001}%
  \BibitemOpen
  \bibfield  {author} {\bibinfo {author} {\bibfnamefont {K.}~\bibnamefont
  {Sengupta}}, \bibinfo {author} {\bibfnamefont {I.}~\bibnamefont {\ifmmode
  \check{Z}\else \v{Z}\fi{}uti\ifmmode~\acute{c}\else \'{c}\fi{}}}, \bibinfo
  {author} {\bibfnamefont {H.-J.}\ \bibnamefont {Kwon}}, \bibinfo {author}
  {\bibfnamefont {V.~M.}\ \bibnamefont {Yakovenko}}, \ and\ \bibinfo {author}
  {\bibfnamefont {S.}~\bibnamefont {Das~Sarma}},\ }\bibfield  {title} {\enquote
  {\bibinfo {title} {Midgap edge states and pairing symmetry of
  quasi-one-dimensional organic superconductors},}\ }\href
  {https://link.aps.org/doi/10.1103/PhysRevB.63.144531} {\bibfield  {journal}
  {\bibinfo  {journal} {Phys. Rev. B}\ }\textbf {\bibinfo {volume} {63}},\
  \bibinfo {pages} {144531} (\bibinfo {year} {2001})}\BibitemShut {NoStop}%
\bibitem [{\citenamefont {Kuerten}\ \emph {et~al.}(2017)\citenamefont
  {Kuerten}, \citenamefont {Richter}, \citenamefont {Mohanta}, \citenamefont
  {Kopp}, \citenamefont {Kampf}, \citenamefont {Mannhart},\ and\ \citenamefont
  {Boschker}}]{Kuerten_PRB2017}%
  \BibitemOpen
  \bibfield  {author} {\bibinfo {author} {\bibfnamefont {L.}~\bibnamefont
  {Kuerten}}, \bibinfo {author} {\bibfnamefont {C.}~\bibnamefont {Richter}},
  \bibinfo {author} {\bibfnamefont {N.}~\bibnamefont {Mohanta}}, \bibinfo
  {author} {\bibfnamefont {T.}~\bibnamefont {Kopp}}, \bibinfo {author}
  {\bibfnamefont {A.}~\bibnamefont {Kampf}}, \bibinfo {author} {\bibfnamefont
  {J.}~\bibnamefont {Mannhart}}, \ and\ \bibinfo {author} {\bibfnamefont
  {H.}~\bibnamefont {Boschker}},\ }\bibfield  {title} {\enquote {\bibinfo
  {title} {In-gap states in superconducting {LaAlO$_3$/SrTiO$_3$} interfaces
  observed by tunneling spectroscopy},}\ }\href
  {https://link.aps.org/doi/10.1103/PhysRevB.96.014513} {\bibfield  {journal}
  {\bibinfo  {journal} {Phys. Rev. B}\ }\textbf {\bibinfo {volume} {96}},\
  \bibinfo {pages} {014513} (\bibinfo {year} {2017})}\BibitemShut {NoStop}%
\bibitem [{\citenamefont {Soumyanarayanan}\ \emph {et~al.}(2017)\citenamefont
  {Soumyanarayanan}, \citenamefont {Raju}, \citenamefont {Gonzalez~Oyarce},
  \citenamefont {Tan}, \citenamefont {Im}, \citenamefont {Petrovi{\'{c}}},
  \citenamefont {Ho}, \citenamefont {Khoo}, \citenamefont {Tran}, \citenamefont
  {Gan}, \citenamefont {Ernult},\ and\ \citenamefont
  {Panagopoulos}}]{Soumyanarayanan_NMat2017}%
  \BibitemOpen
  \bibfield  {author} {\bibinfo {author} {\bibfnamefont {A.}~\bibnamefont
  {Soumyanarayanan}}, \bibinfo {author} {\bibfnamefont {M.}~\bibnamefont
  {Raju}}, \bibinfo {author} {\bibfnamefont {A.~L.}\ \bibnamefont
  {Gonzalez~Oyarce}}, \bibinfo {author} {\bibfnamefont {Anthony K.~C.}\
  \bibnamefont {Tan}}, \bibinfo {author} {\bibfnamefont {M.-Y.}\ \bibnamefont
  {Im}}, \bibinfo {author} {\bibfnamefont {A.~ ~P.}\ \bibnamefont
  {Petrovi{\'{c}}}}, \bibinfo {author} {\bibfnamefont {Pin}\ \bibnamefont
  {Ho}}, \bibinfo {author} {\bibfnamefont {K.~H.}\ \bibnamefont {Khoo}},
  \bibinfo {author} {\bibfnamefont {M.}~\bibnamefont {Tran}}, \bibinfo {author}
  {\bibfnamefont {C.~K.}\ \bibnamefont {Gan}}, \bibinfo {author} {\bibfnamefont
  {F.}~\bibnamefont {Ernult}}, \ and\ \bibinfo {author} {\bibfnamefont
  {C.}~\bibnamefont {Panagopoulos}},\ }\bibfield  {title} {\enquote {\bibinfo
  {title} {Tunable room-temperature magnetic {S}kyrmions in {Ir/Fe/Co/Pt}
  multilayers},}\ }\href {https://doi.org/10.1038/nmat4934} {\bibfield
  {journal} {\bibinfo  {journal} {Nat. Mater.}\ }\textbf {\bibinfo {volume}
  {16}},\ \bibinfo {pages} {898} (\bibinfo {year} {2017})}\BibitemShut
  {NoStop}%
\bibitem [{\citenamefont {Skoropata}\ \emph {et~al.}(2020)\citenamefont
  {Skoropata}, \citenamefont {Nichols}, \citenamefont {Ok}, \citenamefont
  {Chopdekar}, \citenamefont {Choi}, \citenamefont {Rastogi}, \citenamefont
  {Sohn}, \citenamefont {Gao}, \citenamefont {Yoon}, \citenamefont {Farmer},
  \citenamefont {Desautels}, \citenamefont {Choi}, \citenamefont {Haskel},
  \citenamefont {Freeland}, \citenamefont {Okamoto}, \citenamefont {Brahlek},\
  and\ \citenamefont {Lee}}]{Skoropata_SciAdv3902}%
  \BibitemOpen
  \bibfield  {author} {\bibinfo {author} {\bibfnamefont {E.}~\bibnamefont
  {Skoropata}}, \bibinfo {author} {\bibfnamefont {J.}~\bibnamefont {Nichols}},
  \bibinfo {author} {\bibfnamefont {J.~M.}\ \bibnamefont {Ok}}, \bibinfo
  {author} {\bibfnamefont {R.~V.}\ \bibnamefont {Chopdekar}}, \bibinfo {author}
  {\bibfnamefont {E.~S.}\ \bibnamefont {Choi}}, \bibinfo {author}
  {\bibfnamefont {A.}~\bibnamefont {Rastogi}}, \bibinfo {author} {\bibfnamefont
  {C.}~\bibnamefont {Sohn}}, \bibinfo {author} {\bibfnamefont {X.}~\bibnamefont
  {Gao}}, \bibinfo {author} {\bibfnamefont {S.}~\bibnamefont {Yoon}}, \bibinfo
  {author} {\bibfnamefont {T.}~\bibnamefont {Farmer}}, \bibinfo {author}
  {\bibfnamefont {R.~D.}\ \bibnamefont {Desautels}}, \bibinfo {author}
  {\bibfnamefont {Y.}~\bibnamefont {Choi}}, \bibinfo {author} {\bibfnamefont
  {D.}~\bibnamefont {Haskel}}, \bibinfo {author} {\bibfnamefont {J.~W.}\
  \bibnamefont {Freeland}}, \bibinfo {author} {\bibfnamefont {S.}~\bibnamefont
  {Okamoto}}, \bibinfo {author} {\bibfnamefont {M.}~\bibnamefont {Brahlek}}, \
  and\ \bibinfo {author} {\bibfnamefont {H.~N.}\ \bibnamefont {Lee}},\
  }\bibfield  {title} {\enquote {\bibinfo {title} {Interfacial tuning of chiral
  magnetic interactions for large topological {H}all effects in
  {LaMnO$_3$/SrIrO$_3$} heterostructures},}\ }\href
  {https://advances.sciencemag.org/content/6/27/eaaz3902} {\bibfield  {journal}
  {\bibinfo  {journal} {Sci. Adv.}\ }\textbf {\bibinfo {volume} {6}},\ \bibinfo
  {pages} {eaaz3902} (\bibinfo {year} {2020})}\BibitemShut {NoStop}%
\bibitem [{\citenamefont {Lo~Conte}\ \emph {et~al.}(2020)\citenamefont
  {Lo~Conte}, \citenamefont {Nandy}, \citenamefont {Chen}, \citenamefont
  {Fernandes~Cauduro}, \citenamefont {Maity}, \citenamefont {Ophus},
  \citenamefont {Chen}, \citenamefont {N'Diaye}, \citenamefont {Liu},
  \citenamefont {Schmid},\ and\ \citenamefont
  {Wiesendanger}}]{LoConte_NanoLett2020}%
  \BibitemOpen
  \bibfield  {author} {\bibinfo {author} {\bibfnamefont {R.}~\bibnamefont
  {Lo~Conte}}, \bibinfo {author} {\bibfnamefont {A.~K.}\ \bibnamefont {Nandy}},
  \bibinfo {author} {\bibfnamefont {G.}~\bibnamefont {Chen}}, \bibinfo {author}
  {\bibfnamefont {A.~L.}\ \bibnamefont {Fernandes~Cauduro}}, \bibinfo {author}
  {\bibfnamefont {A.}~\bibnamefont {Maity}}, \bibinfo {author} {\bibfnamefont
  {C.}~\bibnamefont {Ophus}}, \bibinfo {author} {\bibfnamefont
  {Z.}~\bibnamefont {Chen}}, \bibinfo {author} {\bibfnamefont {A.~T.}\
  \bibnamefont {N'Diaye}}, \bibinfo {author} {\bibfnamefont {K.}~\bibnamefont
  {Liu}}, \bibinfo {author} {\bibfnamefont {A.~K.}\ \bibnamefont {Schmid}}, \
  and\ \bibinfo {author} {\bibfnamefont {R.}~\bibnamefont {Wiesendanger}},\
  }\bibfield  {title} {\enquote {\bibinfo {title} {Tuning the properties of
  zero-field room temperature ferromagnetic {S}kyrmions by interlayer exchange
  coupling},}\ }\href {https://doi.org/10.1021/acs.nanolett.0c00137} {\bibfield
   {journal} {\bibinfo  {journal} {Nano Lett.}\ }\textbf {\bibinfo {volume}
  {20}},\ \bibinfo {pages} {4739} (\bibinfo {year} {2020})}\BibitemShut
  {NoStop}%
\bibitem [{\citenamefont {Mohanta}\ and\ \citenamefont
  {Taraphder}(2014)}]{Mohanta_EPL2014}%
  \BibitemOpen
  \bibfield  {author} {\bibinfo {author} {\bibfnamefont {N.}~\bibnamefont
  {Mohanta}}\ and\ \bibinfo {author} {\bibfnamefont {A.}~\bibnamefont
  {Taraphder}},\ }\bibfield  {title} {\enquote {\bibinfo {title} {Topological
  superconductivity and {M}ajorana bound states at the {LaAlO$_3$/SrTiO$_3$}
  interface},}\ }\href {https://doi.org/10.1209/0295-5075/108/60001} {\bibfield
   {journal} {\bibinfo  {journal} {{EPL} (Europhysics Letters)}\ }\textbf
  {\bibinfo {volume} {108}},\ \bibinfo {pages} {60001} (\bibinfo {year}
  {2014})}\BibitemShut {NoStop}%
\bibitem [{\citenamefont {Mohanta}\ \emph {et~al.}(2018)\citenamefont
  {Mohanta}, \citenamefont {Kampf},\ and\ \citenamefont
  {Kopp}}]{Mohanta_PRB2018}%
  \BibitemOpen
  \bibfield  {author} {\bibinfo {author} {\bibfnamefont {N.}~\bibnamefont
  {Mohanta}}, \bibinfo {author} {\bibfnamefont {A.~P.}\ \bibnamefont {Kampf}},
  \ and\ \bibinfo {author} {\bibfnamefont {T.}~\bibnamefont {Kopp}},\
  }\bibfield  {title} {\enquote {\bibinfo {title} {Supercurrent as a probe for
  topological superconductivity in magnetic adatom chains},}\ }\href
  {https://link.aps.org/doi/10.1103/PhysRevB.97.214507} {\bibfield  {journal}
  {\bibinfo  {journal} {Phys. Rev. B}\ }\textbf {\bibinfo {volume} {97}},\
  \bibinfo {pages} {214507} (\bibinfo {year} {2018})}\BibitemShut {NoStop}%
\bibitem [{\citenamefont {Tewari}\ and\ \citenamefont
  {Sau}(2012)}]{Tewari_PRL2012}%
  \BibitemOpen
  \bibfield  {author} {\bibinfo {author} {\bibfnamefont {S.}~\bibnamefont
  {Tewari}}\ and\ \bibinfo {author} {\bibfnamefont {J.~D.}\ \bibnamefont
  {Sau}},\ }\bibfield  {title} {\enquote {\bibinfo {title} {Topological
  invariants for spin-orbit coupled superconductor nanowires},}\ }\href
  {https://link.aps.org/doi/10.1103/PhysRevLett.109.150408} {\bibfield
  {journal} {\bibinfo  {journal} {Phys. Rev. Lett.}\ }\textbf {\bibinfo
  {volume} {109}},\ \bibinfo {pages} {150408} (\bibinfo {year}
  {2012})}\BibitemShut {NoStop}%
\end{thebibliography}

\begin{thebibliography}{10}%
\makeatletter
\providecommand \@ifxundefined [1]{%
 \@ifx{#1\undefined}
}%
\providecommand \@ifnum [1]{%
 \ifnum #1\expandafter \@firstoftwo
 \else \expandafter \@secondoftwo
 \fi
}%
\providecommand \@ifx [1]{%
 \ifx #1\expandafter \@firstoftwo
 \else \expandafter \@secondoftwo
 \fi
}%
\providecommand \natexlab [1]{#1}%
\providecommand \enquote  [1]{``#1''}%
\providecommand \bibnamefont  [1]{#1}%
\providecommand \bibfnamefont [1]{#1}%
\providecommand \citenamefont [1]{#1}%
\providecommand \href@noop [0]{\@secondoftwo}%
\providecommand \href [0]{\begingroup \@sanitize@url \@href}%
\providecommand \@href[1]{\@@startlink{#1}\@@href}%
\providecommand \@@href[1]{\endgroup#1\@@endlink}%
\providecommand \@sanitize@url [0]{\catcode `\\12\catcode `\$12\catcode
  `\&12\catcode `\#12\catcode `\^12\catcode `\_12\catcode `\%12\relax}%
\providecommand \@@startlink[1]{}%
\providecommand \@@endlink[0]{}%
\providecommand \url  [0]{\begingroup\@sanitize@url \@url }%
\providecommand \@url [1]{\endgroup\@href {#1}{\urlprefix }}%
\providecommand \urlprefix  [0]{URL }%
\providecommand \Eprint [0]{\href }%
\providecommand \doibase [0]{http://dx.doi.org/}%
\providecommand \selectlanguage [0]{\@gobble}%
\providecommand \bibinfo  [0]{\@secondoftwo}%
\providecommand \bibfield  [0]{\@secondoftwo}%
\providecommand \translation [1]{[#1]}%
\providecommand \BibitemOpen [0]{}%
\providecommand \bibitemStop [0]{}%
\providecommand \bibitemNoStop [0]{.\EOS\space}%
\providecommand \EOS [0]{\spacefactor3000\relax}%
\providecommand \BibitemShut  [1]{\csname bibitem#1\endcsname}%
\let\auto@bib@innerbib\@empty
\bibitem [{\citenamefont {Mohanta}\ \emph
  {et~al.}(2020{\natexlab{a}})\citenamefont {Mohanta}, \citenamefont
  {Okamoto},\ and\ \citenamefont {Dagotto}}]{NM_PRB2020}%
  \BibitemOpen
  \bibfield  {author} {\bibinfo {author} {\bibfnamefont {N.}~\bibnamefont
  {Mohanta}}, \bibinfo {author} {\bibfnamefont {S.}~\bibnamefont {Okamoto}}, \
  and\ \bibinfo {author} {\bibfnamefont {E.}~\bibnamefont {Dagotto}},\
  }\bibfield  {title} {\enquote {\bibinfo {title} {{Planar topological Hall
  effect from conical spin spirals}},}\ }\href
  {https://link.aps.org/doi/10.1103/PhysRevB.102.064430} {\bibfield  {journal}
  {\bibinfo  {journal} {Phys. Rev. B}\ }\textbf {\bibinfo {volume} {102}},\
  \bibinfo {pages} {064430} (\bibinfo {year} {2020}{\natexlab{a}})}\BibitemShut
  {NoStop}%
\bibitem [{\citenamefont {Mohanta}\ \emph
  {et~al.}(2020{\natexlab{b}})\citenamefont {Mohanta}, \citenamefont
  {Christianson}, \citenamefont {Okamoto},\ and\ \citenamefont
  {Dagotto}}]{Mohanta_CommunPhys2020}%
  \BibitemOpen
  \bibfield  {author} {\bibinfo {author} {\bibfnamefont {N.}~\bibnamefont
  {Mohanta}}, \bibinfo {author} {\bibfnamefont {A.~D.}\ \bibnamefont
  {Christianson}}, \bibinfo {author} {\bibfnamefont {S.}~\bibnamefont
  {Okamoto}}, \ and\ \bibinfo {author} {\bibfnamefont {E.}~\bibnamefont
  {Dagotto}},\ }\bibfield  {title} {\enquote {\bibinfo {title} {Signatures of a
  liquid-crystal transition in spin-wave excitations of skyrmions},}\ }\href
  {https://doi.org/10.1038/s42005-020-00489-w} {\bibfield  {journal} {\bibinfo
  {journal} {Commun. Phys.}\ }\textbf {\bibinfo {volume} {3}},\ \bibinfo
  {pages} {229} (\bibinfo {year} {2020}{\natexlab{b}})}\BibitemShut {NoStop}%
\bibitem [{\citenamefont {Mohanta}\ and\ \citenamefont
  {Taraphder}(2014)}]{Mohanta_EPL2014}%
  \BibitemOpen
  \bibfield  {author} {\bibinfo {author} {\bibfnamefont {N.}~\bibnamefont
  {Mohanta}}\ and\ \bibinfo {author} {\bibfnamefont {A.}~\bibnamefont
  {Taraphder}},\ }\bibfield  {title} {\enquote {\bibinfo {title} {Topological
  superconductivity and {M}ajorana bound states at the {LaAlO$_3$/SrTiO$_3$}
  interface},}\ }\href {https://doi.org/10.1209/0295-5075/108/60001} {\bibfield
   {journal} {\bibinfo  {journal} {{EPL} (Europhysics Letters)}\ }\textbf
  {\bibinfo {volume} {108}},\ \bibinfo {pages} {60001} (\bibinfo {year}
  {2014})}\BibitemShut {NoStop}%
\bibitem [{\citenamefont {Mohanta}\ \emph {et~al.}(2018)\citenamefont
  {Mohanta}, \citenamefont {Kampf},\ and\ \citenamefont
  {Kopp}}]{Mohanta_PRB2018}%
  \BibitemOpen
  \bibfield  {author} {\bibinfo {author} {\bibfnamefont {N.}~\bibnamefont
  {Mohanta}}, \bibinfo {author} {\bibfnamefont {A.~P.}\ \bibnamefont {Kampf}},
  \ and\ \bibinfo {author} {\bibfnamefont {T.}~\bibnamefont {Kopp}},\
  }\bibfield  {title} {\enquote {\bibinfo {title} {Supercurrent as a probe for
  topological superconductivity in magnetic adatom chains},}\ }\href
  {https://link.aps.org/doi/10.1103/PhysRevB.97.214507} {\bibfield  {journal}
  {\bibinfo  {journal} {Phys. Rev. B}\ }\textbf {\bibinfo {volume} {97}},\
  \bibinfo {pages} {214507} (\bibinfo {year} {2018})}\BibitemShut {NoStop}%
\bibitem [{\citenamefont {Sticlet}\ \emph {et~al.}(2012)\citenamefont
  {Sticlet}, \citenamefont {Bena},\ and\ \citenamefont
  {Simon}}]{Sticlet_PRL2012}%
  \BibitemOpen
  \bibfield  {author} {\bibinfo {author} {\bibfnamefont {D.}~\bibnamefont
  {Sticlet}}, \bibinfo {author} {\bibfnamefont {C.}~\bibnamefont {Bena}}, \
  and\ \bibinfo {author} {\bibfnamefont {P.}~\bibnamefont {Simon}},\ }\bibfield
   {title} {\enquote {\bibinfo {title} {Spin and {M}ajorana polarization in
  topological superconducting wires},}\ }\href
  {https://link.aps.org/doi/10.1103/PhysRevLett.108.096802} {\bibfield
  {journal} {\bibinfo  {journal} {Phys. Rev. Lett.}\ }\textbf {\bibinfo
  {volume} {108}},\ \bibinfo {pages} {096802} (\bibinfo {year}
  {2012})}\BibitemShut {NoStop}%
\bibitem [{\citenamefont {Sedlmayr}\ and\ \citenamefont
  {Bena}(2015)}]{Sedlmayr_PRB2015}%
  \BibitemOpen
  \bibfield  {author} {\bibinfo {author} {\bibfnamefont {N.}~\bibnamefont
  {Sedlmayr}}\ and\ \bibinfo {author} {\bibfnamefont {C.}~\bibnamefont
  {Bena}},\ }\bibfield  {title} {\enquote {\bibinfo {title} {Visualizing
  {M}ajorana bound states in one and two dimensions using the generalized
  {M}ajorana polarization},}\ }\href
  {https://link.aps.org/doi/10.1103/PhysRevB.92.115115} {\bibfield  {journal}
  {\bibinfo  {journal} {Phys. Rev. B}\ }\textbf {\bibinfo {volume} {92}},\
  \bibinfo {pages} {115115} (\bibinfo {year} {2015})}\BibitemShut {NoStop}%
\bibitem [{\citenamefont {Scharf}\ \emph {et~al.}(2019)\citenamefont {Scharf},
  \citenamefont {Pientka}, \citenamefont {Ren}, \citenamefont {Yacoby},\ and\
  \citenamefont {Hankiewicz}}]{Scharf_PRB2019}%
  \BibitemOpen
  \bibfield  {author} {\bibinfo {author} {\bibfnamefont {B.}~\bibnamefont
  {Scharf}}, \bibinfo {author} {\bibfnamefont {F.}~\bibnamefont {Pientka}},
  \bibinfo {author} {\bibfnamefont {H.}~\bibnamefont {Ren}}, \bibinfo {author}
  {\bibfnamefont {A.}~\bibnamefont {Yacoby}}, \ and\ \bibinfo {author}
  {\bibfnamefont {E.~M.}\ \bibnamefont {Hankiewicz}},\ }\bibfield  {title}
  {\enquote {\bibinfo {title} {Tuning topological superconductivity in
  phase-controlled {J}osephson junctions with {R}ashba and {D}resselhaus
  spin-orbit coupling},}\ }\href
  {https://link.aps.org/doi/10.1103/PhysRevB.99.214503} {\bibfield  {journal}
  {\bibinfo  {journal} {Phys. Rev. B}\ }\textbf {\bibinfo {volume} {99}},\
  \bibinfo {pages} {214503} (\bibinfo {year} {2019})}\BibitemShut {NoStop}%
\bibitem [{\citenamefont {Sengupta}\ \emph {et~al.}(2001)\citenamefont
  {Sengupta}, \citenamefont {\ifmmode \check{Z}\else
  \v{Z}\fi{}uti\ifmmode~\acute{c}\else \'{c}\fi{}}, \citenamefont {Kwon},
  \citenamefont {Yakovenko},\ and\ \citenamefont
  {Das~Sarma}}]{Sengupta_PRB2001}%
  \BibitemOpen
  \bibfield  {author} {\bibinfo {author} {\bibfnamefont {K.}~\bibnamefont
  {Sengupta}}, \bibinfo {author} {\bibfnamefont {I.}~\bibnamefont {\ifmmode
  \check{Z}\else \v{Z}\fi{}uti\ifmmode~\acute{c}\else \'{c}\fi{}}}, \bibinfo
  {author} {\bibfnamefont {H.-J.}\ \bibnamefont {Kwon}}, \bibinfo {author}
  {\bibfnamefont {V.~M.}\ \bibnamefont {Yakovenko}}, \ and\ \bibinfo {author}
  {\bibfnamefont {S.}~\bibnamefont {Das~Sarma}},\ }\bibfield  {title} {\enquote
  {\bibinfo {title} {Midgap edge states and pairing symmetry of
  quasi-one-dimensional organic superconductors},}\ }\href
  {https://link.aps.org/doi/10.1103/PhysRevB.63.144531} {\bibfield  {journal}
  {\bibinfo  {journal} {Phys. Rev. B}\ }\textbf {\bibinfo {volume} {63}},\
  \bibinfo {pages} {144531} (\bibinfo {year} {2001})}\BibitemShut {NoStop}%
\bibitem [{\citenamefont {Kuerten}\ \emph {et~al.}(2017)\citenamefont
  {Kuerten}, \citenamefont {Richter}, \citenamefont {Mohanta}, \citenamefont
  {Kopp}, \citenamefont {Kampf}, \citenamefont {Mannhart},\ and\ \citenamefont
  {Boschker}}]{Kuerten_PRB2017}%
  \BibitemOpen
  \bibfield  {author} {\bibinfo {author} {\bibfnamefont {L.}~\bibnamefont
  {Kuerten}}, \bibinfo {author} {\bibfnamefont {C.}~\bibnamefont {Richter}},
  \bibinfo {author} {\bibfnamefont {N.}~\bibnamefont {Mohanta}}, \bibinfo
  {author} {\bibfnamefont {T.}~\bibnamefont {Kopp}}, \bibinfo {author}
  {\bibfnamefont {A.}~\bibnamefont {Kampf}}, \bibinfo {author} {\bibfnamefont
  {J.}~\bibnamefont {Mannhart}}, \ and\ \bibinfo {author} {\bibfnamefont
  {H.}~\bibnamefont {Boschker}},\ }\bibfield  {title} {\enquote {\bibinfo
  {title} {In-gap states in superconducting {LaAlO$_3$/SrTiO$_3$} interfaces
  observed by tunneling spectroscopy},}\ }\href
  {https://link.aps.org/doi/10.1103/PhysRevB.96.014513} {\bibfield  {journal}
  {\bibinfo  {journal} {Phys. Rev. B}\ }\textbf {\bibinfo {volume} {96}},\
  \bibinfo {pages} {014513} (\bibinfo {year} {2017})}\BibitemShut {NoStop}%
\end{thebibliography}

%




\pagebreak
\setcounter{equation}{0}
\setcounter{figure}{0}
\setcounter{table}{0}
\setcounter{page}{1}
\makeatletter
\renewcommand{\theequation}{E\arabic{equation}}
\renewcommand{\thefigure}{S\arabic{figure}}

\noindent \textbf{\large Supplementary Information}\\

\begin{figure}[t]
\vspace{0mm}
\begin{center}
\epsfig{file=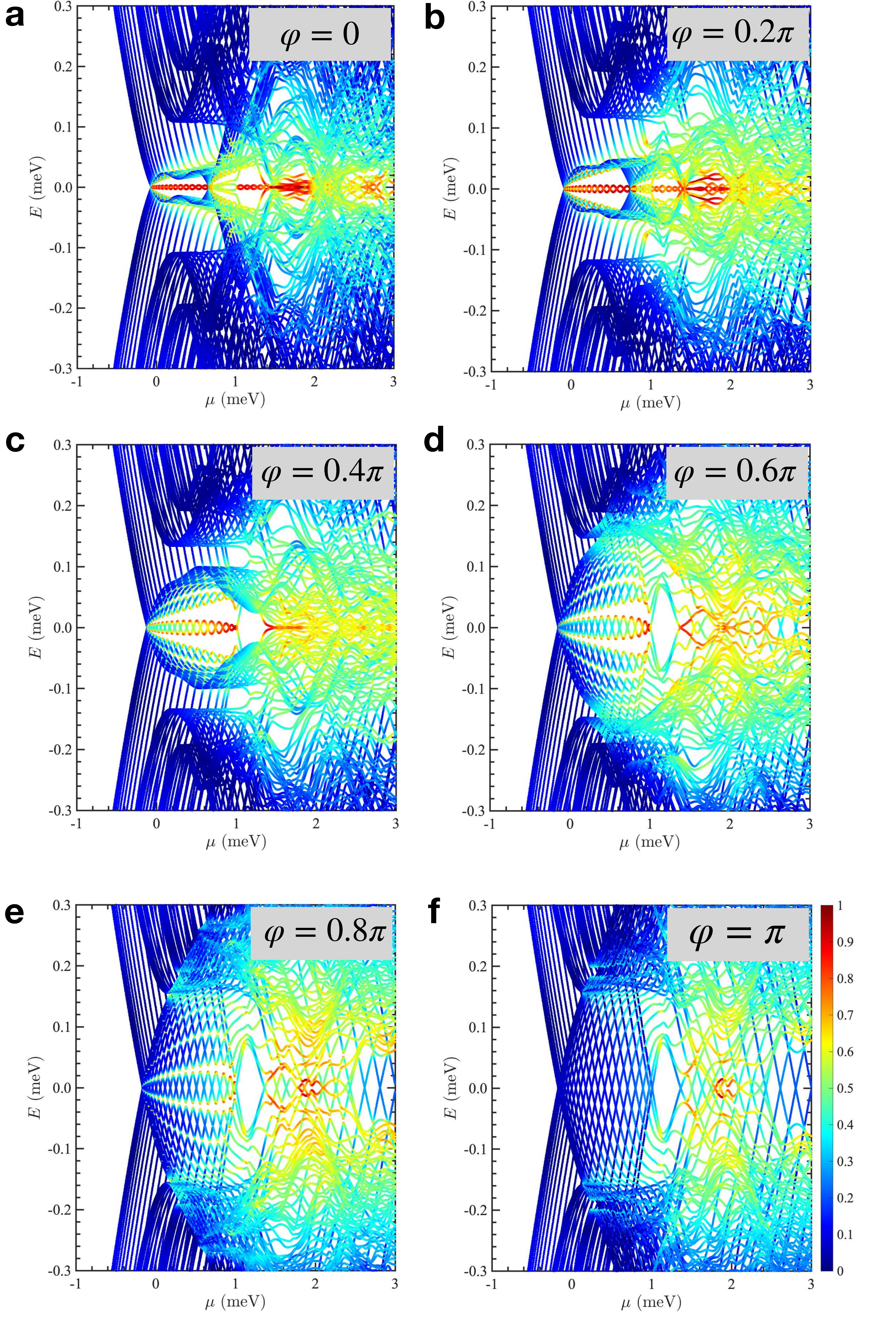,trim=0.0in 0.1in 0.0in 0.0in,clip=true, width=84mm}
\caption{{\bf Majorana oscillation in quasiparticle spectrum.} Quasiparticle spectrum of a planar Josephson junction, attached to a skyrmion crystal as considered in Fig.2 of the main text, with varying chemical potential $\mu$ at different values of the phase difference $\varphi$ between the two superconducting regions of the Josephson junction: {\bf a} $\varphi=0$, {\bf b} $\varphi=0.2\pi$, {\bf c} $\varphi=0.4\pi$, {\bf d} $\varphi=0.6\pi$, {\bf e} $\varphi=0.8\pi$, and {\bf f} $\varphi=\pi$. The colorbar represents the Majorana polarization $|{\cal P}_{{\cal M},1}|$ of the lowest positive eigenstate. The intrinsic Rashba spin-orbit coupling is set to zero. All other parameters are as in the Fig.2 of the main text.}
\label{figS1}
\vspace{-8mm}
\end{center}
\end{figure}

\begin{figure}[t]
\begin{center}
\epsfig{file=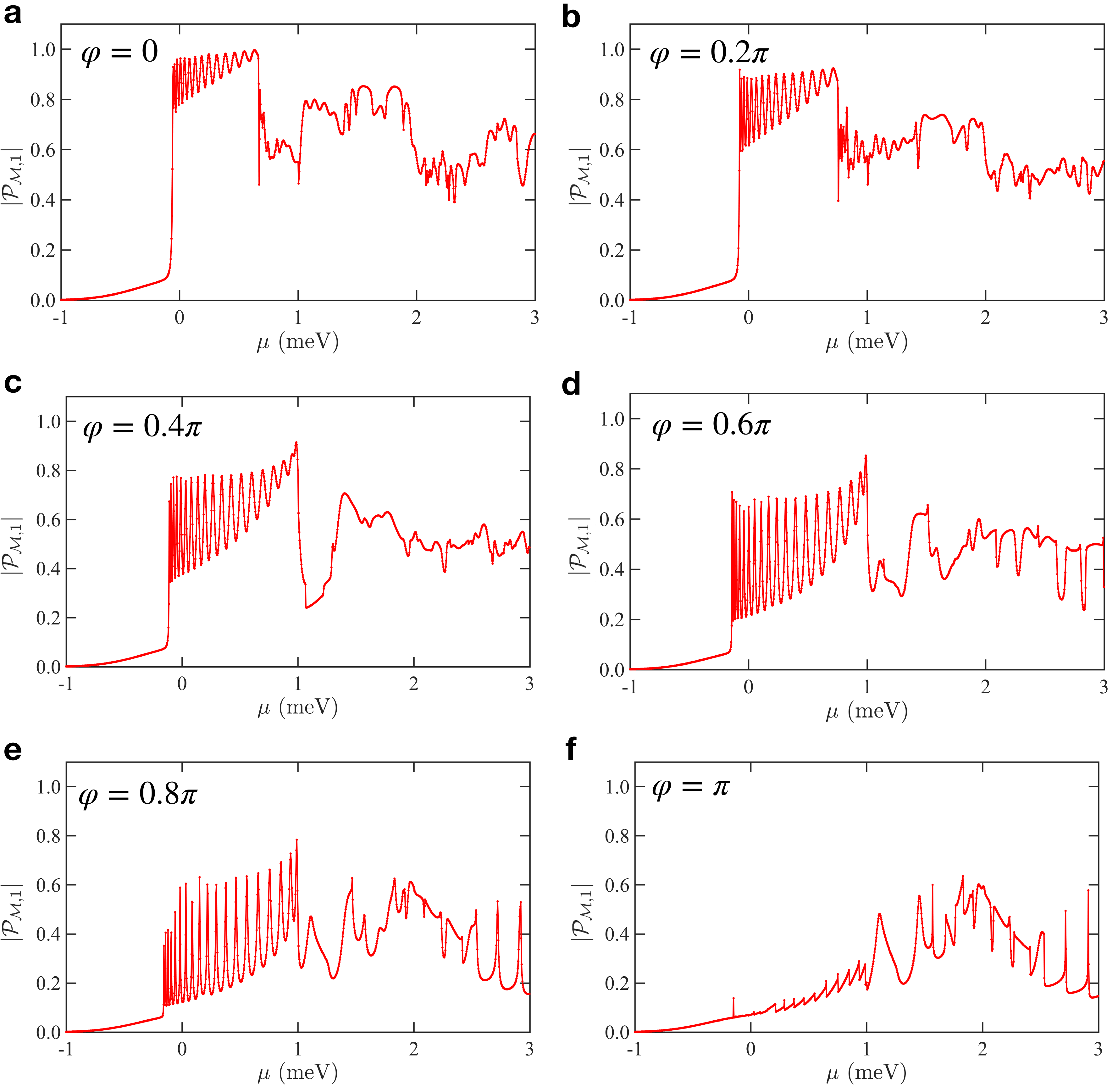,trim=0.0in 0.0in 0.0in 0.0in,clip=false, width=85mm}
\caption{{\bf Majorana oscillation in Majorana polarization.} The Majorana polarization $|{\cal P}_{{\cal M},1}|$ of the lowest positive energy eigenstate with varying chemical potential $\mu$ at different values of the phase difference $\varphi$: {\bf a} $\varphi=0$, {\bf b} $\varphi=0.2\pi$, {\bf c} $\varphi=0.4\pi$, {\bf d} $\varphi=0.6\pi$, {\bf e} $\varphi=0.8\pi$, and {\bf f} $\varphi=\pi$. The intrinsic Rashba spin-orbit coupling is kept at zero. All other parameters are as in the Fig.2 of the main text.}
\label{figS2}
\vspace{-6mm}
\end{center}
\end{figure}

\vspace{2em}
\noindent \small{\bf Supplementary Note 1: Majorana oscillation in the Josephson junctions}\\
\noindent The oscillation of the Majorana bound states (MBS) is a well-known phenomenon that appears due to the finite size effect. In a one-dimensional geometry, \textit{e.g.} a Rashba spin-orbit coupled nanowire with a Zeeman exchange coupling, the probability density of the MBS, which are localized dominantly at the two ends, decay exponentially with distance towards the middle of the nanowire. The overlap of these two MBS wave functions give rise to a finite splitting in energy. This split energy gap between the MBS pair oscillates with varying a parameter, such as the chemical potential or the Zeeman energy. In our considered planar Josephson junction geometry, the finite length of the quasi-one-dimensional metallic channel in the middle, therefore, naturally leads to the oscillations of the zero-energy MBS with varying chemical potential $\mu$. Furthermore, the finite width of the quasi-one-dimensional channel provides extra room for delocalization of the MBS at the two ends, contributing additively to the Majorana oscillation. The oscillation amplitude, however, decreases with increasing the length of the Josephson junction. In Fig.~\ref{figS1}, we show the quasiparticle spectra with varying $\mu$ at different values of the phase difference $\varphi$ between the two superconducting regions of the Josephson junction. Evidently, the oscillation amplitude increases with increasing $\varphi$ from 0 to $\pi$. At $\varphi \!=\! \pi$, the oscillation becomes large and the MBS completely vanish. The oscillation can also be visible in the Majorana polarization $|{\cal P}_{{\cal M},1}|$ of the lowest positive eigenstate. In Fig.~\ref{figS2}, we show the variation in $|{\cal P}_{{\cal M},1}|$ for the lowest positive eigenstate with $\mu$. At $\varphi \!=\! 0$, the oscillation  is small in amplitude and appears on the top of a significant value of the Majorana polarization ($|{\cal P}_{{\cal M},1}|$$\approx$1). With increasing $\varphi$ from 0 to $\pi$, the value of $|{\cal P}_{{\cal M},1}|$ decreases and the oscillation amplitude increases simultaneously. These results explain why the MBS disappear with increasing the phase difference $\varphi$ from $0$ to $\pi$, as discussed in the main text.\\

\begin{figure}[t]
\vspace{0mm}
\begin{center}
\epsfig{file=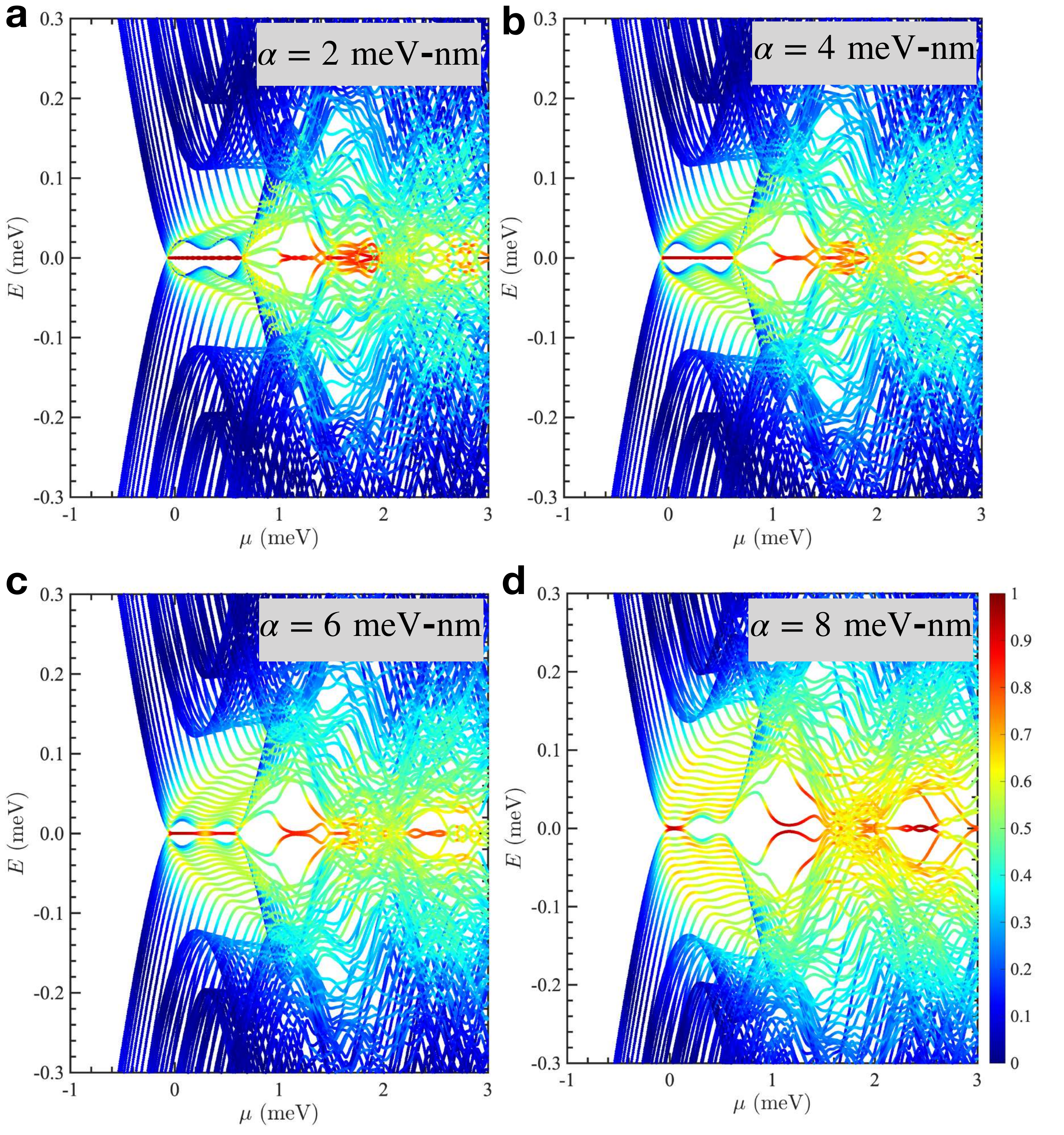,trim=0.0in 0.0in 0.0in 0.0in,clip=false, width=88mm}
\caption{{\bf Influence of Rashba spin-orbit coupling.} Quasiparticle spectrum of the planar Josephson junction as considered in Fig.2 of the main text at phase difference $\varphi \!=\!0$ with varying chemical potential at different values of the intrinsic Rashba spin-orbit coupling: {\bf a} $\alpha \!=\!2$~meV-nm, {\bf b} $\alpha \!=\!4$~meV-nm, {\bf c} $\alpha \!=\!6$~meV-nm, and {\bf d} $\alpha \!=\!8$~meV-nm. The colorbar represents the Majorana polarization $|{\cal P}_{{\cal M},1}|$ of the lowest positive eigenstate. The Majorana bound states remain robust in the presence of the intrinsic Rashba spin-orbit coupling.}
\label{figS3}
\vspace{-8mm}
\end{center}
\end{figure}

\noindent \small{\bf Supplementary Note 2: Effect of Rashba spin-orbit coupling}\\
\noindent The broken inversion symmetry at the interface between the two-dimensional electron gas and the superconductor, often leads to a sizable intrinsic Rashba spin-orbit coupling which is usually considered as the primary mechanism for modifying the pairing symmetry of the induced superconductivity in a one-dimensional~\cite{Mohanta_EPL2014} or two-dimensional geometry~\cite{Mohanta_PRB2018}, leading to the desired topological superconductivity. To explore the mutual influence of the Rashba spin-orbit coupling and the Skyrmion crystal (SkX)-generated spin-orbit coupling on the emergence of the zero-energy MBS, we consider the Hamiltonian ${\cal H}_{\rm RSOC}$=$-i\alpha/(2a)\sum_{\langle ij \rangle,\sigma,\sigma^{\prime}}(\boldsymbol{\sigma}\! \times \!{\hat{r}_{ij}})_{\sigma \sigma^{\prime}}^{z}c_{i\sigma}^{\dagger}c_{j\sigma^{\prime}}$ for the Rashba spin-orbit coupling, where $\alpha$ is the strength of the Rashba spin-orbit coupling, and obtain the quasiparticle spectrum based on the total Hamiltonian ${\cal H}_{\rm BdG}+{\cal H}_{\rm RSOC}$. 
In Fig.~\ref{figS3}, we show the quasiparticle spectra for four different values of $\alpha$. With relatively smaller values of $\alpha$,  \textit{e.g.} {\bf a.} $\alpha \!=\!2$~meV-nm and {\bf b.} $\alpha \!=\!4$~meV-nm, the zero-energy MBS within the range -0.2 meV~$\lessapprox \mu \lessapprox$~0.6 meV, remain stable. The oscillations of the MBS are minimized within this range of $\mu$, making the MBS more robust (\textit{i.e.} localized). For larger values of $\alpha$ (cases {\bf c.} and {\bf d.}), the MBS in the above range of $\mu$ start to become delocalized, possibly because a very large Rashba spin-orbit coupling interferes destructively with the synthetic spin-orbit coupling. But remarkably, with larger values of $\alpha$, new MBS start to appear at larger values of the chemical potential, \textit{e.g.} within the range 2.2 meV~$\lessapprox \mu \lessapprox$~2.6 meV. Although the mutual effects of the intrinsic Rashba spin-orbit coupling and the SkX-generated spin-orbit coupling are rather complex, within some ranges of $\mu$, they cooperate to generate robust MBS with a larger topological energy gap.\\

\begin{figure*}[t]
\vspace{0mm}
\begin{center}
\epsfig{file=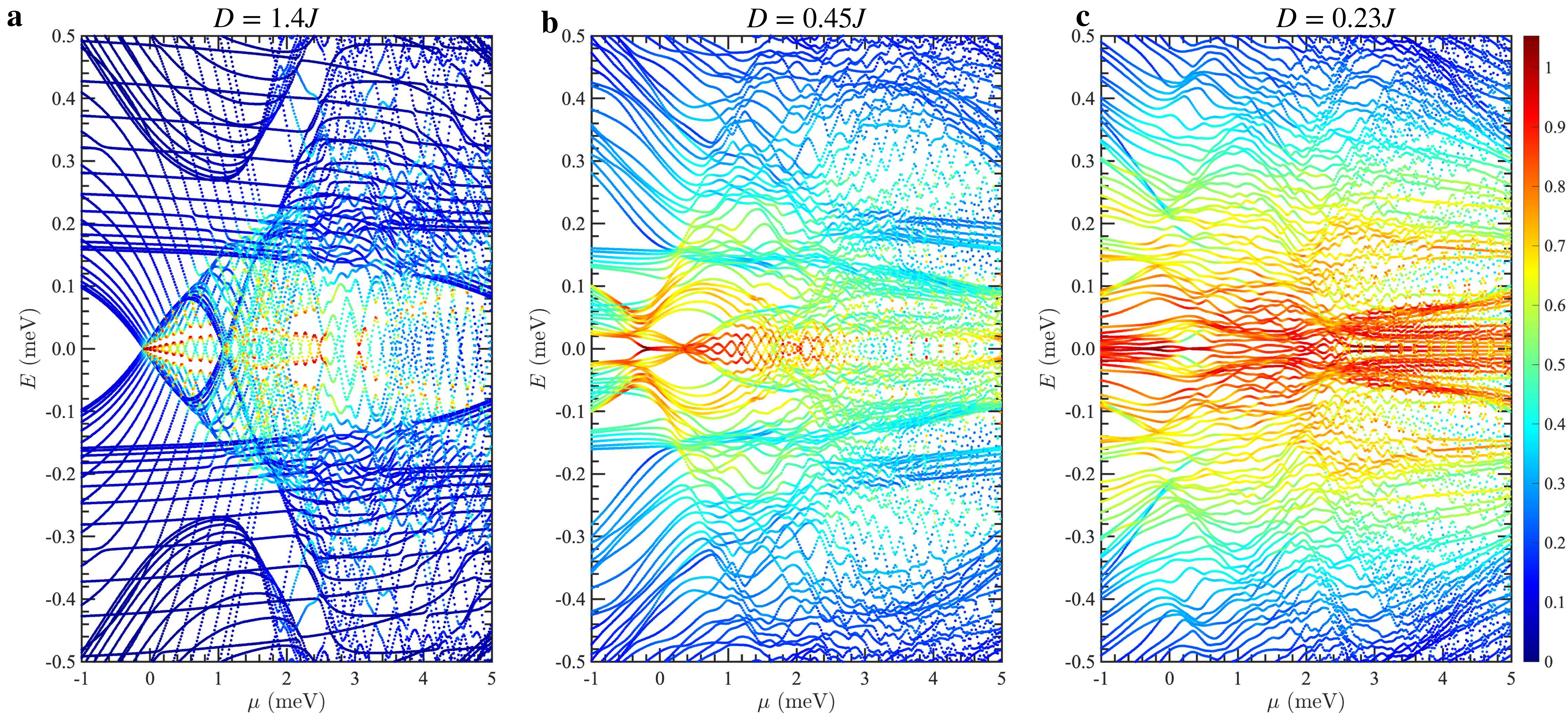,trim=0.0in 0.0in 0.0in 0.0in,clip=false, width=165mm}
\caption{{\bf Quasiparticle spectrum with skyrmion crystal only below the quasi-1D channel.} Quasiparticle spectrum, with varying chemical potential $\mu$, and three different values of the Dzyaloshinskii-Moriya interaction strength {\bf a} $D=1.4J$, {\bf b} $D=0.45J$, and {\bf c} $D=0.23J$. The colorbar represents the Majorana polarization $|{\cal P}_{{\cal M},1}|$ of the lowest positive eigenstate. All other parameters are the same as in Fig. 2 of the main text.}
\label{figS4}
\vspace{-7mm}
\end{center}
\end{figure*}

\begin{figure}[htb!]
\vspace{0mm}
\begin{center}
\epsfig{file=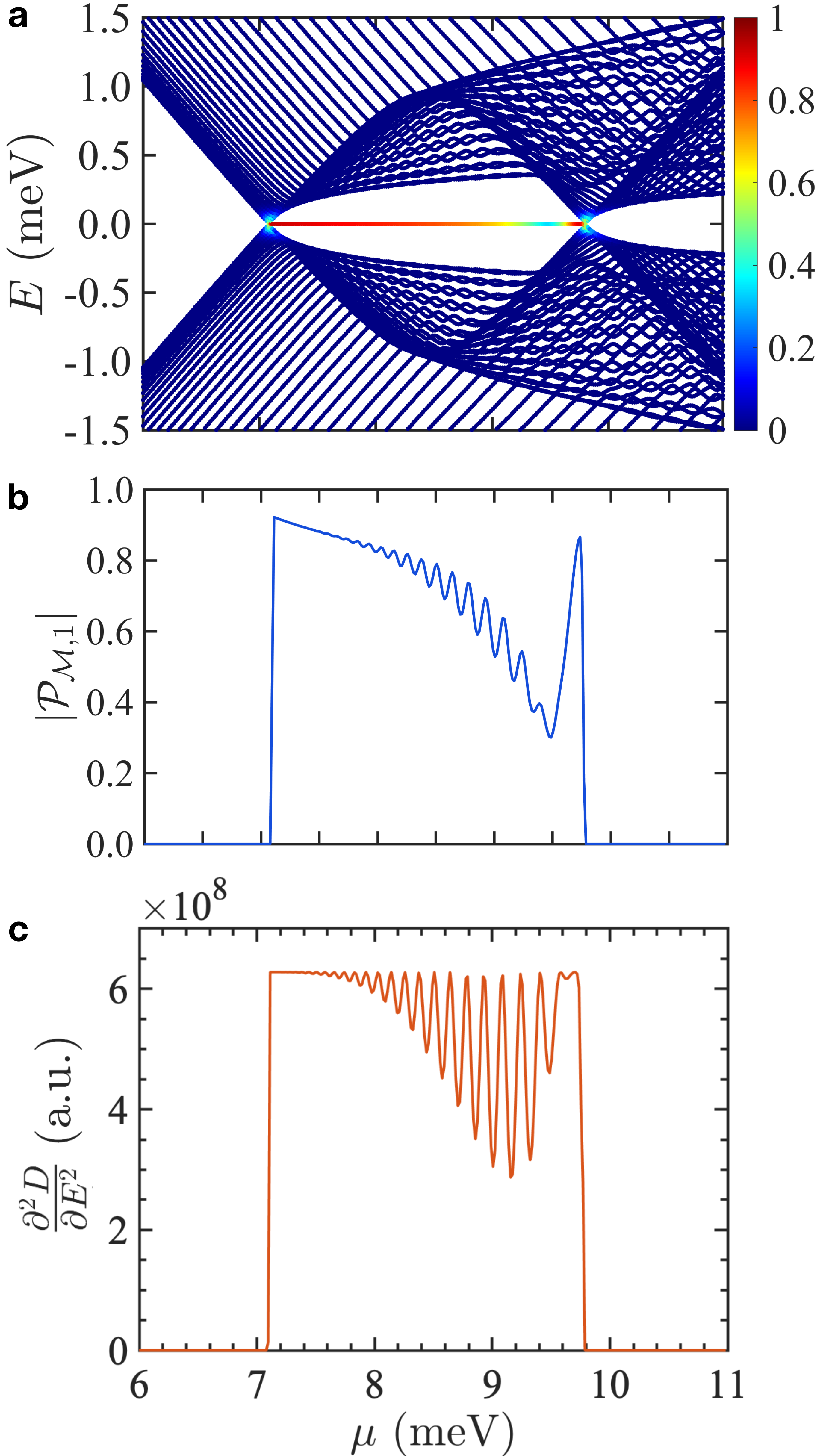,trim=0.0in 0.0in 0.0in 0.0in,clip=false, width=46mm}
\caption{{\bf Order parameters in the wire geometry.} {\bf a} Quasiparticle spectrum, with varying chemical potential $\mu$, of a wire of length $L_x \!=\!190a$ ($a$ being the unit lattice spacing) that is placed at the middle of a superconductor of size $200a$$\times$$5a$. A Zeeman field $B\!=\!0.3$~T is uniformly present throughout the wire. The Rashba spin-orbit coupling strength is $\alpha \!=\!30$~meV-nm. The g-factor is $g\!=\!200$. The superconducting potential is $U\!=\!2$~meV and the pairing amplitude $\Delta_i$ is treated self-consistently. The colorbar represents the Majorana polarization $|{\cal P}_{{\cal M},n}|$ of the eigenstate with energy $\epsilon_n$. {\bf b} The $\mu$-variation of $|{\cal P}_{{\cal M},1}|$. {\bf c} The $\mu$-variation of  $\frac{\partial^2 D}{\partial E^2}$, the curvature of the density of states at zero energy.}
\label{figS5}
\vspace{-7mm}
\end{center}
\end{figure}

\noindent \small{\bf Supplementary Note 3: Skyrmion crystal below the quasi-1D channel}\\ 
\noindent We studied the possibility of obtaining the MBS with the SkX, lying only below the non-superconducting region of the planar Josephson junction. As shown by the spectra in Fig.~\ref{figS4} for three different values of the DMI strength, there is no signature of the formation of robust MBS with strongly localized nature. The overall influence of the skyrmion spin texture on the quasiparticle spectrum is lesser than the case when the SkX is placed below the entire Josephson junction. We would further like to stress on the fact that the planar Josephson junction operates in a slightly different mechanism than the 1D nanowire setting. In the 1D nanowire set up, the topological superconductivity is realized within the nanowire, while in the planar Josephson junction, the MBS appear in the middle region, a region that is void of superconducting pairing. Therefore, the influence of the skyrmion texture on the superconducting regions is also important to realize the strongly localized MBS in the non-superconducting channel. \\

\noindent \small{\bf Supplementary Note 4: Order parameters for the Majorana bound states}\\
\noindent From the solution of the Bogoliubov-de Gennes equations of the planar Josephson junction with the realistic SkX texture, we identify the MBS in the quasiparticle spectrum and track the transition to the topological superconducting phase by computing two quantities: (i) the generalized Majorana polarization $\cal{P_M}$, proposed in Refs.~\cite{Sticlet_PRL2012,Sedlmayr_PRB2015}, and (ii) the second derivative of the total density of states at zero energy,  $\frac{\partial^2D}{\partial E^2}$, also called as the curvature of the density of states. The second derivative of the local density of states at a boundary site and at zero energy was used as an order parameter in Ref.~\cite{Scharf_PRB2019} to track the topological superconducting transition. These quantities may provide additional insight to detect the MBS in experiments, besides  the conventional zero-bias conductance peak~\cite{Sengupta_PRB2001,Elbio_arxiv2020} which often leads to ambiguity due to other possible zero-bias states in a superconductor~\cite{Kuerten_PRB2017}. To test the above two quantities in our set up, we first consider a chain with Rashba spin-orbit coupling and a uniform Zeeman exchange coupling, attached on the top and at the middle of a quasi-one-dimensional $s$-wave superconductor. The BdG quasiparticle spectrum of such a system, shown in Fig.~\ref{figS5}{\bf a}, depicts the emergence of the zero-energy MBS within a range of the chemical potential $\mu$. The $\mu$-variation of the two above-discussed quantities, $|{\cal P}_{{\cal M},1}|$ and $\frac{\partial^2D}{\partial E^2}$, are plotted in, respectively, Fig.~\ref{figS5}{\bf b} and {\bf c}. Interestingly, both these quantities reveal sharp jumps on entering the topological superconducting phase, similar to an order parameter in an ordinary phase transition. The Majorana oscillations, although not noticeable in the quasiparticle spectrum, is visible in both $|{\cal P}_{{\cal M},1}|$ and $\frac{\partial^2D}{\partial E^2}$. In the infinite-length limit, the Majorana oscillations vanish and both these quantities reveal a quantized feature, as shown previously in Ref.~\onlinecite{Sticlet_PRL2012}.
 The sharp jumps, therefore, attest that these two quantities can reliably be used to track the topological superconducting transition in our computational framework.\\

\end{document}